\documentclass[a4,aps,amsmath,floatfix,nofootinbib,longbibliography,twocolumn,prb,superscriptaddress]{revtex4-1} 

\usepackage{graphicx} 
\usepackage{amsmath}
\usepackage{enumerate} 
\usepackage{braket}
\usepackage[caption=false]{subfig}
\usepackage[colorlinks=true,linkcolor=red]{hyperref}
\usepackage{xcolor}

\newcommand{\tG}{\tilde{G}}
\newcommand{\tg}{\tilde{g}}

\newcommand{\xioe}{\xi_1^{(e)}}

\newcommand{\pn}{\text{PN}}

\newcommand{\mh}{\mathcal{H}}
\newcommand{\tpn}{\text{PN}}
\newcommand{\emx}{E_\text{max}}
\newcommand{\mD}{\mathcal{D}}

\newcommand{\pcsadd}{Center for Theoretical Physics of Complex Systems, Institute for Basic Science (IBS), Daejeon, Korea, 34126}
\newcommand{\ustadd}{Basic Science Program, Korea University of Science and Technology (UST), Daejeon, Korea, 34113}

\graphicspath{{./Figures/}}
\begin{document}

\title{Multifractality of correlated two-particle bound states in quasiperiodic chains}

\author{Diana Thongjaomayum}
\affiliation{\pcsadd}

\author{Sergej Flach}
\affiliation{\pcsadd}
\affiliation{\ustadd}

\author{Alexei Andreanov}
\affiliation{\pcsadd}
\affiliation{\ustadd}

\date{\today}

\begin{abstract}
    We consider the quasiperiodic Aubry-Andr\'e chain in the insulating regime with localised single-particle states. Adding local interaction leads to the emergence of extended correlated two-particle bound states. We analyse the nature of these states including their multifractality properties. We use a projected Green function method to compute numerically participation numbers of eigenstates and analyse their dependence on the energy and the system size. We then perform a scaling analysis. We observe multifractality of correlated extended two-particle bound states, which we confirm independently through exact diagonalisation. 
\end{abstract}


\maketitle

\section{Introduction}

Understanding the transport properties of quantum disordered or inhomogeneous systems has been an active topic of research since the discovery of Anderson localisation (AL). AL describes the arrest of transport in a single particle system due to disorder or inhomogeneous potential which renders all the eigenstates in one and two space dimensions exponentially localised.~\cite{anderson1958absence} The original work of Anderson triggered a sequence of theoretical studies and by now the single particle case is well understood.~\cite{kramer1993localization} The important and much harder question is the stability of modification of AL in the presence of many-body interactions.  Decades of research attempts culminated in the opening of the field of Many Body Localisation.\cite{basko2006metal,abanin2019colloquium,fabien2018many} Interestingly one of the strongly debated issues is the possible existence of 'bad' metallic states which are non-ergodic or simply multi-fractal.~\cite{basko2006metal,fabien2018many}

A notorious issue with MBL related studies is the computational complexity due to the exponential proliferation of the Hilbert space dimension with increasing numbers of particles and system size. A legitimate and complementary approach is therefore to consider only few interacting particles, which allows to increase the system size beyond the limits set by typical MBL models. Three main directions with single particle localisation in one dimension have been explored: genuine AL due to uncorrelated disorder,~\cite{kramer1993localization} Wannier-Stark (WSL) localisation due to an external dc field,~\cite{fukuyama1973tightly} and Aubry-Andr\'e (AAL) localisation due to a quasiperiodic external potential.~\cite{aubry1980analyticity} Genuine AL yields a nontrivial increase of the localisation length for two interacting particles with still unsettled scaling details.~\cite{shepelyansky1994coherent,frahm1999interaction,krimer2011two,thongjaomayum2019taming} Two interacting particles yield no localisation change for WSL with interaction, only affecting the Bloch oscillation  periods.~\cite{khomeriki2010interaction} At variance, AAL with quasiperiodic potentials showed an unexpected transition from localisation (zero interaction) to delocalisation (non-zero interaction).~\cite{flach2012correlated} These findings were later confirmed in Ref.~\onlinecite{frahm2015freed} which provided additional indications for the fractal nature of the delocalised eigenstates. 

Are these the seeds of a bad metal and the MBL transition from above? A hint might be obtained from the striking similarity of the phase diagram of correlated metallic two-particle bound states in Fig.~4 of Ref.~\onlinecite{flach2012correlated} and the phase diagram  of an MBL phase which was experimentally assessed for interacting fermions in optical quasiperiodic potentials in Fig.~4 of Schreiber \textit{et al} in Ref.~\onlinecite{schreiber2015observation}. In the present study we attempt to add more conclusive arguments which aim at a positive answer for the above question for quasiperiodic potentials.  We confirm the fractal character of the two-particle spectrum and the fractalilty of some of the two-particle states. We rely on the projected Green function method,~\cite{vonoppen1996interaction} originally developed to analyse the localisation length of two interacting particles in the AL case. The paper is organised as follows: we introduce the tools and other necessary means in Sec.~\ref{sec:stage}. Section~\ref{sec:sp} benchmarks these tools in the single particle case against the exact results and exact diagonalisation. In Sec.~\ref{sec:tip} we analyse the two interacting particles case. This is followed by conclusions.

\section{Setting the stage}
\label{sec:stage}

The starting point is a single particle placed in a quasiperiodic potential with the Aubry-Andr\'e Hamiltonian~\cite{aubry1980analyticity}
\begin{gather}
    \label{eq:ham-1p}
    \mh_0 = \sum_n (\vert n\rangle \langle n+1 + \text{h.c}) + \sum_m h_m,\\
    h_n = \lambda\cos(2\pi\alpha n + \beta),\notag
\end{gather}
where $\lambda$ is the strength of the potential, $\alpha$ is an irrational number ensuring quasiperiodicity of the potential. We choose $\alpha=(\sqrt{5} - 1)/2$, the golden ratio and we fix the hopping strength $t=1$. Depending on the strength of the potential $\lambda$ the eigenstates are all delocalised ($\lambda < 2$) or localised ($\lambda > 2$) with localisation length $\xi_1=1/\ln(\lambda/2)$, which is the same for all the eigenstates.~\cite{aubry1980analyticity} Finally $\beta$ is a phase which can be varied to generate different realisations of the quasiperiodic potential. In numerical studies with finite system size the choice of $\beta$ will affect localised and sparse, fractal or multi-fractal extended states. In the present study involving critical states we use averaging over different values of $\beta$, that we denote as $\overline{\cdots}$, to improve statistics.

We now add the interactions and consider two interacting bosons. We choose the onsite Hubbard interaction of strength $u$. The total Hamiltonian is given  by
\begin{multline}
    \label{eq:ham-2p}
    \mh = \mh_0\otimes\mh_0 + uP\\
    = \sum_{n,m}({|n,m\rangle\langle n+1,m| + |n,m\rangle\langle n,m+1| + \text{h.c.})}\\
    + \sum_{n,m}|n,m\rangle(h_n + h_m)\langle n,m| + u P,
\end{multline}
where $|n,m\rangle$ is a basis state with two particles at site $n,m$, and $h_n$ is the onsite Aubry-Andr\'e potential at site $n$ given by Eq.~\eqref{eq:ham-1p}. $P$ is the projection operator defined as $P|n,m\rangle=\delta_{nm}|n,m\rangle$ that enforces the onsite Hubbard interaction.

The authors of the work Ref.~\onlinecite{flach2012correlated} used exact diagonalisation and unitary evolution of wavepackets to study the two-particle properties of the model~\eqref{eq:ham-2p}. The exact diagonalisation limited the largest system sizes achievable to $N\approx 250$, imposed by the efficiency of full diagonalisation of the Hamiltonian matrix~\eqref{eq:ham-2p}. Later Frahm in Ref.~\onlinecite{frahm2015freed} implemented a dedicated sparse diagonalisation algorithm based on Green functions~\cite{arnoldi1951the,vonoppen1996interaction} to handle large sizes, up to $N=10946$, of the Hamiltonian~\eqref{eq:ham-2p}. We follow the original approach of Ref.~\onlinecite{vonoppen1996interaction}. We extract the relevant two-particles properties from the projected two-particle Green function, which is obtained as a projection of the full Green function $G=(E-\mh)^{-1}$ onto doubly occupied states (relying crucially on the fact that the Hubbard interaction is proportional to the projector $P$): 
\begin{gather}
    \tG = \frac{\tG_0}{1 - u\tG_0}.
\end{gather}
Here $\tG = PGP$ and $\tG_0=PG_0P$; $G_0$ is the non-interacting two particle GF which can be obtained by straightforward
diagonalisation of the single particle Hamiltonian~\eqref{eq:ham-1p}. Knowing the single particle eigenenergies $\{E_\mu\}$ and eigenfunctions $\{\phi_{\mu}(n)\}$ we compute $G_0$ as follows:
\begin{gather}
    \langle n,n| G_0(E)|m,m\rangle = \sum_{\mu,\nu} \frac{\phi_\mu(n)\phi_\nu(n)\phi_\mu^*(m)\phi_\nu^*(m)}{E - E_\mu - E_\nu}\notag\\
    \label{eq:G0}
    = \sum_\mu \phi_\mu(n)g_0(E - E_\mu)\phi_\mu(m),\\
    g_0(E) = \frac{1}{E - \mh_0} = \sum_\nu \frac{\phi_\nu^*(n)\phi_\nu(m)}{E - E_\nu}.\notag
\end{gather}
The reordering of the terms in the second line is done to reduce the complexity of the computation from the original $O(N^4)$ to $O(N^3)$,~\cite{frahm1999interaction} since the single particle Green function $g_0$ can be efficiently evaluated using tridiagonal matrix inversion of the single particle Hamiltonian~\eqref{eq:ham-1p}. This approach allows us achieve system sizes as large as $N=7000$.

In the insulating regime the exponential decay of the projected Green's function $\tG$ was used to extract the two interacting particles (TIP) localisation length.~\cite{vonoppen1996interaction,frahm2016eigenfunction,thongjaomayum2019taming} Here we are aiming to investigate TIP eigenstates which we expect to be extended in a predominantly insulating region,~\cite{flach2012correlated} therefore $\tG$ might not decay or the decay might not be exponential.
Consequently we adopt a different measure:~\cite{thongjaomayum2019taming} Interpreting the projected Green function $\tG$ as a probability density function we define the participation number $I_{q=2}$ and its higher moments $I_{q>2}$ as
\begin{gather}
    \label{eq:I2-def}
    I_q = (\sum_k|\tg(k)|)^q/\sum_k|\tg(k)|^q,
\end{gather} 
where $\tg(k)=<n,n|\tG|n+k,n+k>$. We shall use $I_2$ and higher moments that are always well defined to analyse the TIP states. To distinguish $I_q$ from the conventional participation number we will refer to it as the Green function participation number (GPN). However before we can proceed to the two particle case, we need to confirm that $I_2$ is a valid measure of localisation of an eigenstate $\Psi$, similar to the conventional participation number:
\begin{gather}
    \tpn_q = \sum_{nm}\vert \Psi_{nm}\vert^{2q},
\end{gather}
e.g. that $I_2$ can distinguish between extended, (multi)fractal and localised states.

\section{Single particle: benchmarking}
\label{sec:sp}

\begin{figure}[htb]
    \subfloat[]{
        \includegraphics[width=0.45\textwidth]{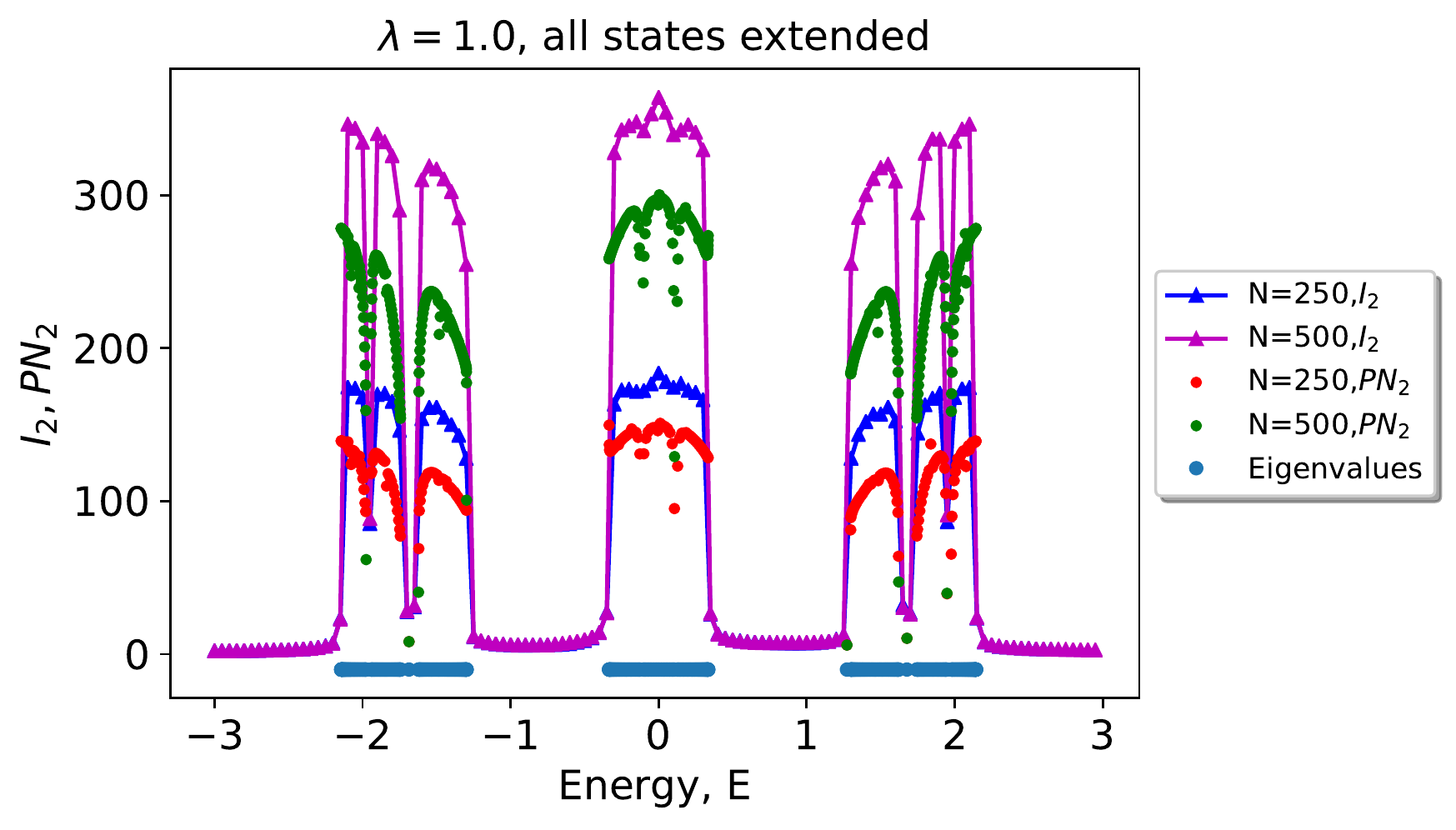}}
        
    \subfloat[]{
        \includegraphics[width=0.45\textwidth]{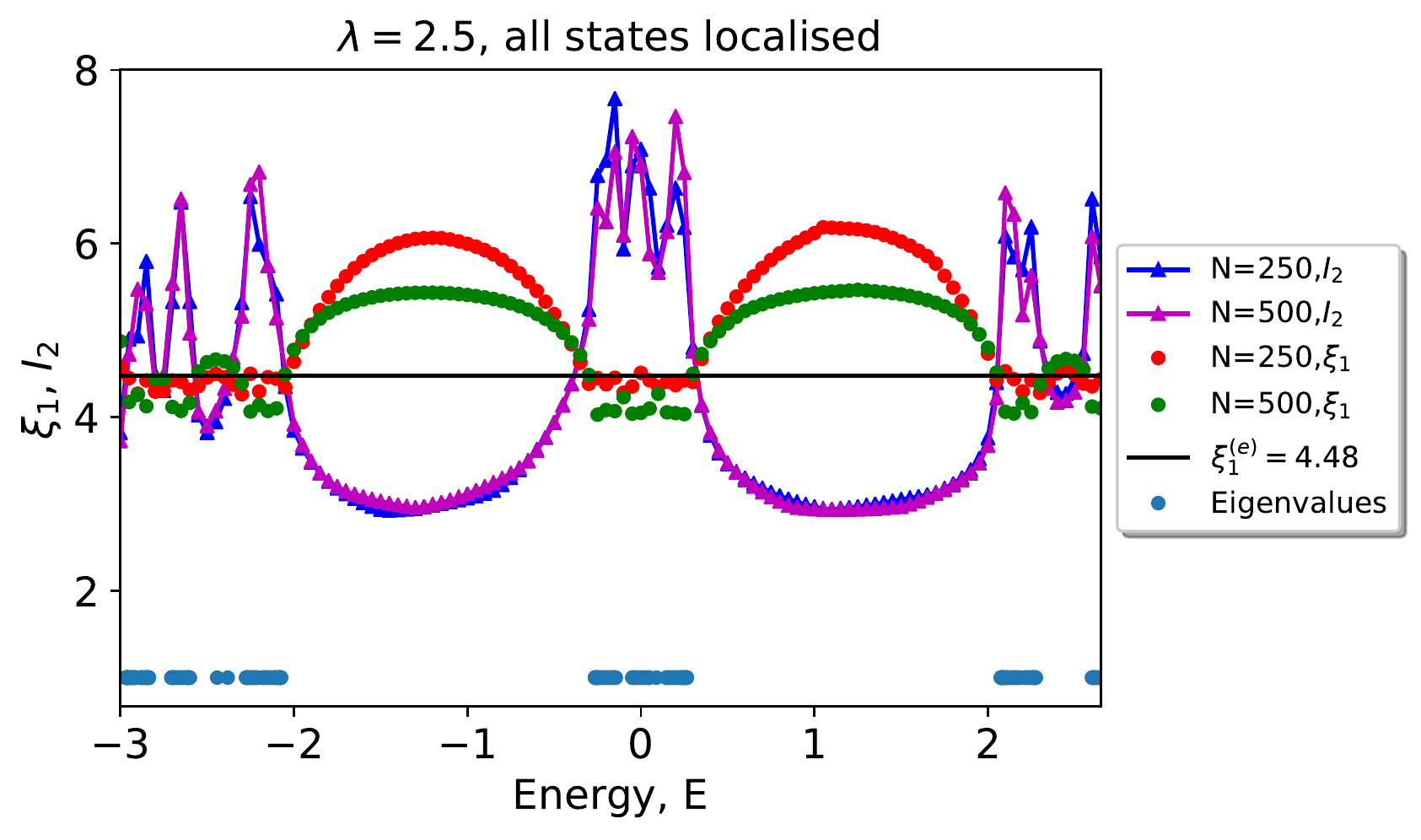}}
    \caption{(Colour online) Benchmarking of a single particle in the AA model: (a) $\lambda=1$: $I_2$ and $\tpn_2$ for $N=250, 500$. (b) $\lambda=2.5$: Localisation length $\xi_1$ and $I_2$ for $N=250,500$. The black line represents the analytical value $\xi_1=4.48$. The participation number $\tpn_2$ behaves similarly to $I_2$ (not shown). The bottom blue/circular points in both (a) and (b) represent the spectrum of $\mh_0$ for $N=500$ and show the locations of the eigenstates.}
    \label{fig:1}
\end{figure}

To confirm that the above defined participation number $I_q$ is a valid probe of localisation properties of eigenstates  we first consider the single particle case. To achieve this we benchmark two single-particle quantities: localisation length -- analytical and numerical
\begin{gather}
    \xi_1 = \frac{1}{\ln\left(\frac{\lambda}{2}\right)}\\
    \label{eq:xi1-def}
    \frac{1}{\xi_1} = -\lim_{|n-m| \to\infty} \frac{\overline{\ln|\langle n|g_0 |m\rangle|}}{|n-m|},
\end{gather} 
and participation number $I_2$, which is defined similarly to its two particle version Eq.~\eqref{eq:I2-def}:
\begin{gather}
    \label{eq:I21-def}
    I_q = (\sum_k|g(k)|)^q/\sum_k|g(k)|^q,
\end{gather} 
where $g(k)=<n|g_0|n+k>$. Our aim is to confirm that $I_2$ is a valid substitute for $\xi_1$ for localised states and it behaves like the conventional participation number for localised and extended states.

To prove that we consider $3$ different values of the potential strength $\lambda=1,2,2.5$, which correspond to  delocalised, critical and localised regimes. For each $\lambda$ we compute single particle eigenstates $\psi$ and the Green function $g_0$ for energies $E \in [-3,3]$ in steps of $\Delta E=0.05$, scanning the entire single particle spectrum. This step size of $0.05$ is chosen to be slightly bigger than the level spacing $\delta(N=250)=0.004$ and $\delta(N=500)=0.002$ for the data presented in Fig.~\ref{fig:1}. From the Green function $g_0$ we evaluate $\xi_1$ and $I_2$ and from the eigenstates $\psi$ we compute $\tpn_2$. Figure~\ref{fig:1}(a) shows the results for $\lambda=1$ for which all the single particle eigenstates are extended. The plot of Fig.~\ref{fig:1}(a) shows $I_2$ and $\tpn_2$ vs $E$ for system sizes $N=250,500$. The bottom circular/blue points of Fig.~\ref{fig:1}(a) show the single-particle spectrum obtained from the full diagonalisation of $\mh_0$ for $N=500$. We see that both participation numbers $\tpn_2$ and $I_2$ drop to zero in in the gaps of the spectrum of $\mh_0$, and increase with the system size for energies where eigenstates are present. We observe $I_2>\tpn_2$ in general.~\cite{thongjaomayum2019taming} Figure~\ref{fig:1}(b) compares the same quantities for $\lambda=2.5$ where the entire spectrum is localised. Fig.~\ref{fig:1}(b) shows $I_2$ and $\xi_1$ against $E$ for $N=250,500$. The eigenenergies are plotted at the bottom of Fig.~\ref{fig:1}(b) (light blue points). The black line is the exact localisation length $\xioe=1/\ln(1.25)\approx 4.48$. The localisation length $\xi_1$  evaluated from the Green function~\eqref{eq:xi1-def} is close to the exact value $\xioe$ for energies close to the eigenenergies of the system, while $I_2$ is systematically larger than $\xi_1$, but is roughly of the same order, and does not scale with the system size $N$. In the gaps of the exact spectrum, $I_2$ drops to zero which is expected since there are no eigenstates corresponding to these energies and contributions from the eigenstates are negligible. However $\xi_1$, defined by Eq.~\eqref{eq:xi1-def}, gives completely wrong value in the gaps of the single particle spectrum as seen in Fig.~\ref{fig:1}(b). This is clearly an artefact of the exponential fitting of $g_0$ that does not decay exponentially inside the gaps of the spectrum of $\mh_0$. The behaviour of the participation number $\tpn_2$  is very similar to $I_2$ (not shown). For the critical case $\lambda=2$, the behaviour of $\xi_1$ and $I_2$ is similar to that of the delocalised $\lambda=1$ case.

 \begin{figure}[htb]
    \centering
    \includegraphics[width=0.45\textwidth,angle=0]{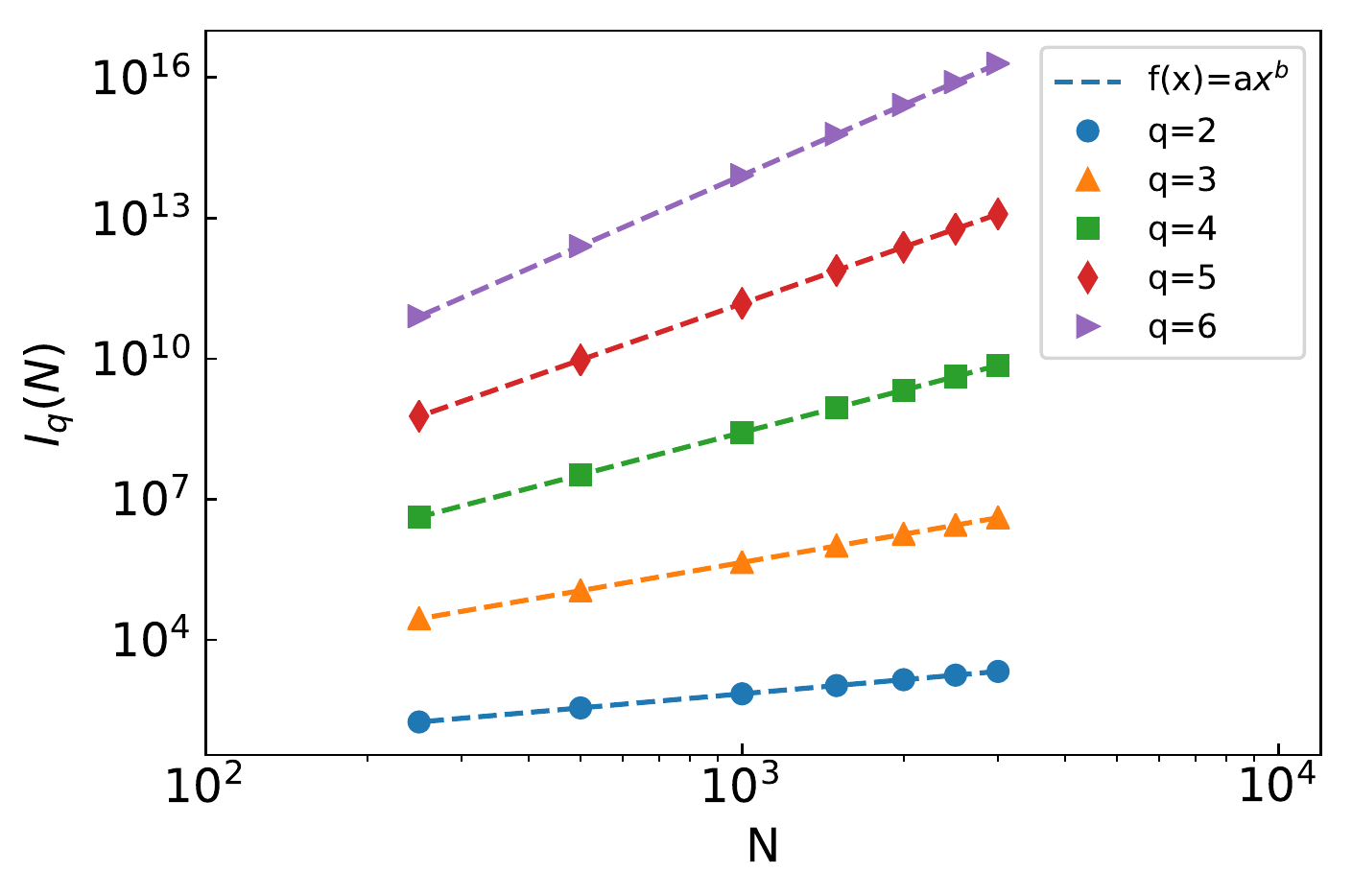}
    \caption{(Colour online) Participation number $I_q(N)$ (symbols) vs system size $N=250$ to $3000$ for $q=2,3,4,5,6$ and the power law fits $I_q(N)=a N^b$ (dashed lines) for $\lambda=1$. The power-law fit works well also for $\lambda=2,2.5$ (not shown).}
    \label{fig:2}
\end{figure}

 \begin{figure}[htb]
    \centering
    \includegraphics[width=0.45\textwidth,angle=0]{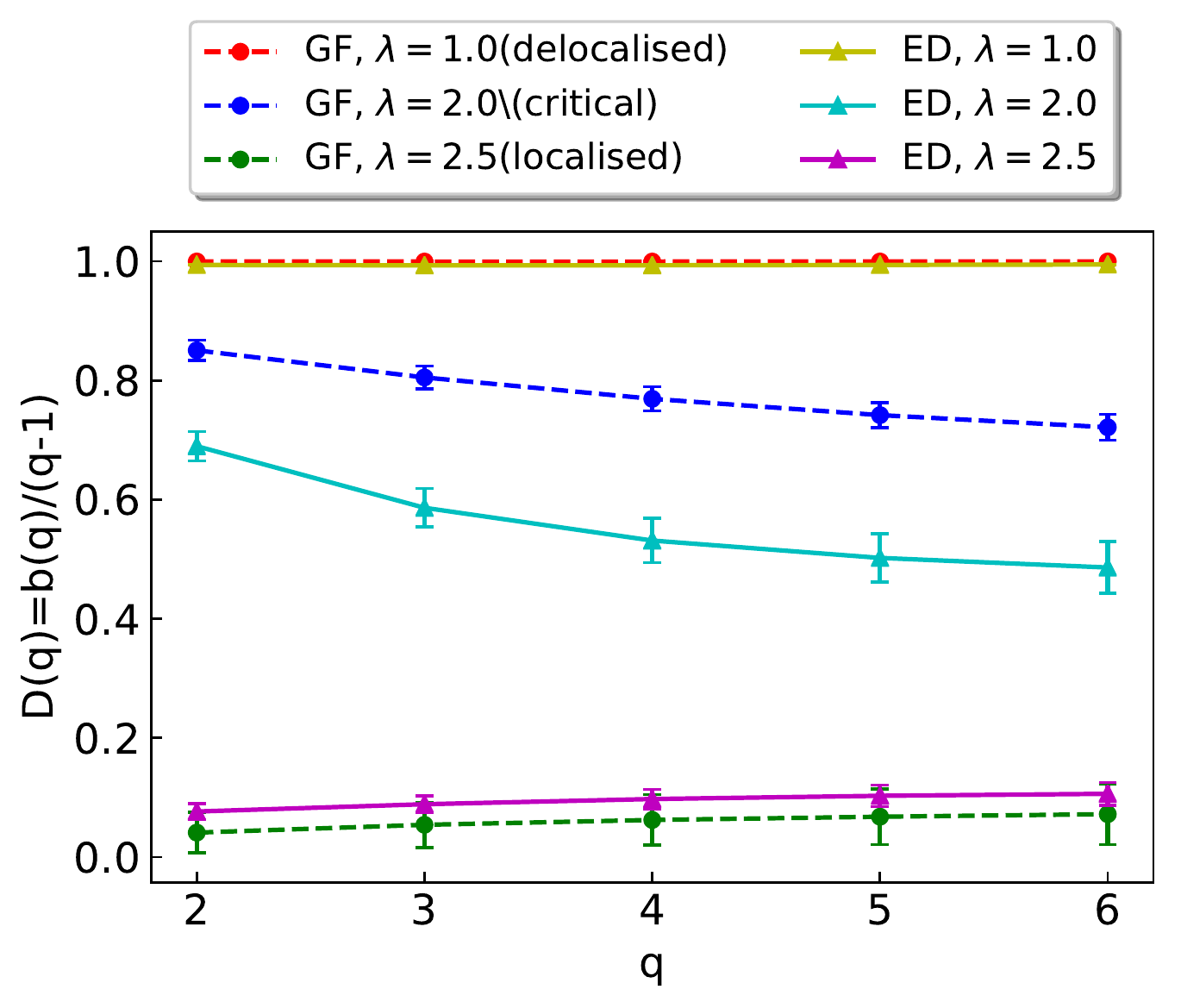}
    \caption{(Colour online) Fractal dimension $D_q$ vs $q$ obtained from Green's function (GF) and exact diagonalisation (ED) in extended (red/yellow), critical (blue/cyan) and localised (green/magenta) regimes for the single particle case. The dimension $D_q$ is $q$-independent and equal to zero(one) in the extended(localised) regime and has a non-trivial dependence on $q$ at the criticality, $\lambda=2$. This a similar behaviour to the dimension $\mD_q$ computed from the $\tpn_q$.}
    \label{fig:3}
\end{figure}

This rough comparison lends support to the validity of $I_2$ as a substitute for the participation number $\tpn_2$. To strengthen this support we look into the scaling of the participation numbers $\tpn_q$ with the power $q$, which also distinguishes extended, localised and (multi)fractal states: $\tpn_q = a N^{\mD_q(q-1)}$ where $\mD_q$ is the fractal dimension of the state, and $\mD_q=0$ corresponds to localised state, $\mD_q=1$ corresponds to delocalised states, $0<\mD_q<1$ - to (multi)fractal states. We verify whether a similar scaling holds for $I_q$ and try the fit $I_q=a N^{D_q(q-1)}$ for all the three regimes: $\lambda=1, 2, 2.5$. We pick the energy $\emx$ corresponding to the maximum of $I_2$ for the largest system size considered, $N = 3000$, since we want to probe the most delocalised states in an otherwise localised regime (this choice is only relevant for $\lambda=2.5$ where all eigenstates are localised), and use this value $\emx$ to evaluate $D_q$ for smaller system sizes. For every $\lambda$ we compute $I_q$ for $q=2,3,4,5,6$ and for a range of system sizes $N=250$ to $3000$. The results are shown in Fig.~\ref{fig:2}: we see a clear power law scaling of $I_q$ with $N$ for every individual value of $q$. Next we fit these data for several system sizes to extract $D_q$ for the values of $\lambda=1,2,2.5$. Similarly we evaluate the $\mD_q$ from the scaling of $\tpn_q$ with system size. The $\tpn_q$ are computed from exact diagonalisation of a single particle Hamiltonian~\eqref{eq:ham-1p}. The results are summarised in Fig.~\ref{fig:3}: both methods agree - $\mD_q\approx D_q\approx 1$ for $\lambda=1.0$ - as it should be for extended states, $\mD_q\approx D_q\approx 0.0$ - for the localised case $\lambda=2.5$, and $q$-dependent $\mD_q$, $D_q$ for the critical value $\lambda=2.0$ where multifractality is expected.

These results indicate that $I_2$ can be used as a substitute for the localisation length $\xi_1$ and the participation number $\pn_q$ in the single particle case. We assume that this is also the case for two interacting particles and verify this assumption self-consistently. Therefore in what follows we will study the behaviour of $I_2$ and higher moments $I_{q>2}$.

\section{Two interacting particles: self-similarity of the spectrum and fractality of the eigenstates}
\label{sec:tip}

We now turn to the case of two particles with the onsite Hubbard interaction. Earlier work~\cite{flach2012correlated} has reported the emergence of metallic states in the single-particle insulating regime ($\lambda>2$). This conclusion was based on exact diagonalisation of systems up to $N=250$ (up to $N=1000$ with sparse diagonalisation) sites and analysis of the spreading of time-evolved wave packets scanned in the entire range of interactions $0<u<12$ for several values of potential strength $\lambda\in [1.8,3]$. These results were enhanced by Frahm~\cite{frahm2015freed}, who performed diagonalisation of systems up to $N=10946$ sites and confirmed the presence of delocalised states.

\begin{figure}[htb]
    \centering
    \includegraphics[width=0.4\textwidth,angle=0]{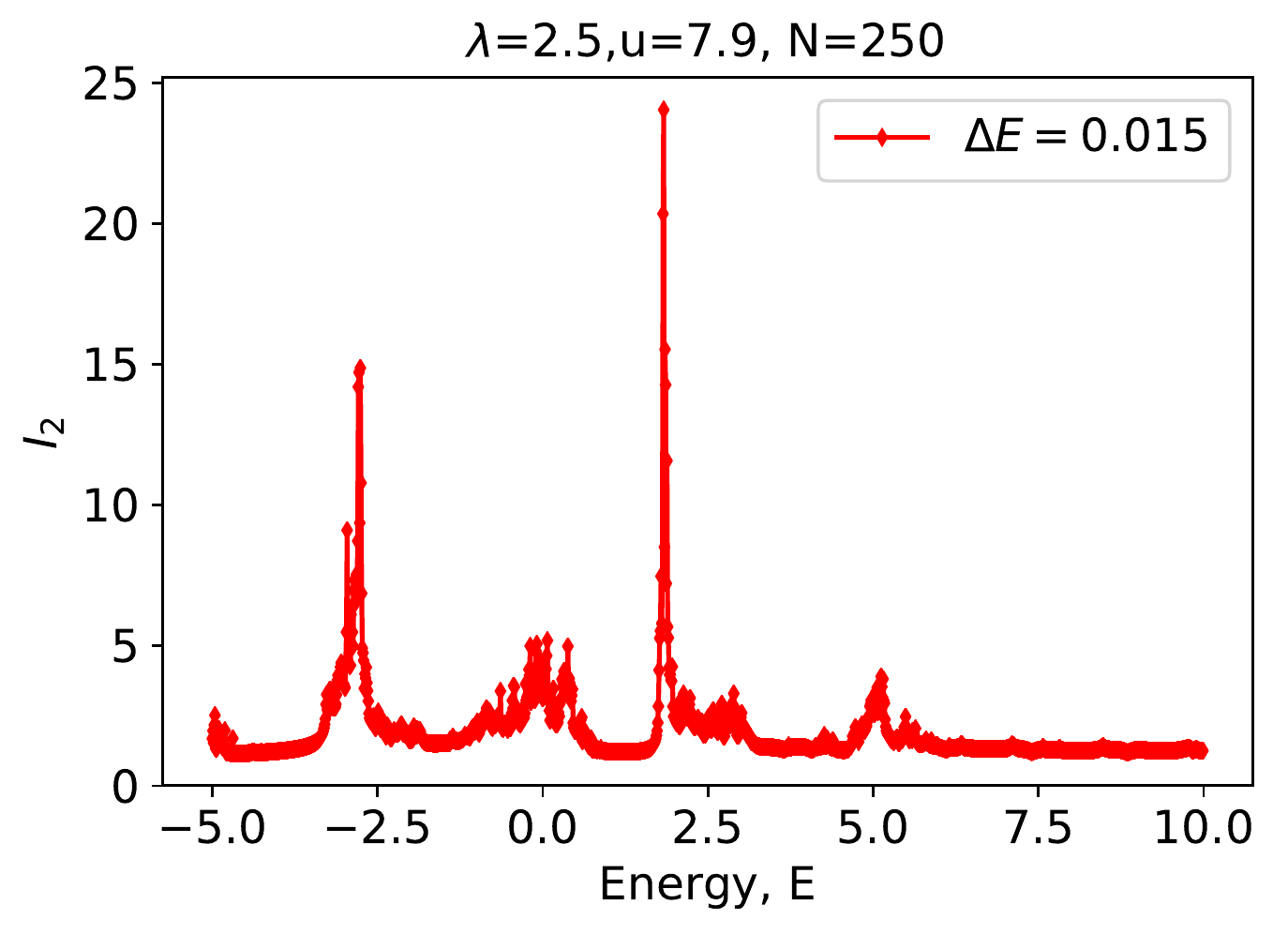}
    \caption{(Colour online) Participation number $I_2$ vs energy $E$ for two interacting particles. The peaks signal the emergence of delocalised states in the otherwise localised spectrum.}
    \label{fig:4}
\end{figure}

We start our analysis with a cross check of Ref.~\onlinecite{flach2012correlated} and evaluate the participation number $I_2$ from $\tG_{nm}$~\eqref{eq:I2-def} for $\lambda=2.5$, $u=7.9$ and $1000$ values of energy $E\in [-5,10]$. The results are averaged over $10$ disorder realisations, e.g. values of $\beta$, see Eq.~\eqref{eq:ham-1p}. In Fig.~\ref{fig:4} we see a minibands structure with the few energies where the value of $I_2$ is relatively large, similarly to the findings of Ref.~\onlinecite{flach2012correlated} thereby lending further support to the use of $I_2$ as a probe of the extent of the eigenstates. We identified two values of energy, $E_1\approx 1.8, E_2\approx -2.8$ where $I_2$ achieves its local maximum (Fig.~\ref{fig:4}), suggesting the emergence of delocalised states at these energies.

\begin{figure*}[tbh]
    \centering{}\subfloat[]{
    \begin{centering}
        \includegraphics[width=0.24\textwidth,height=3cm]{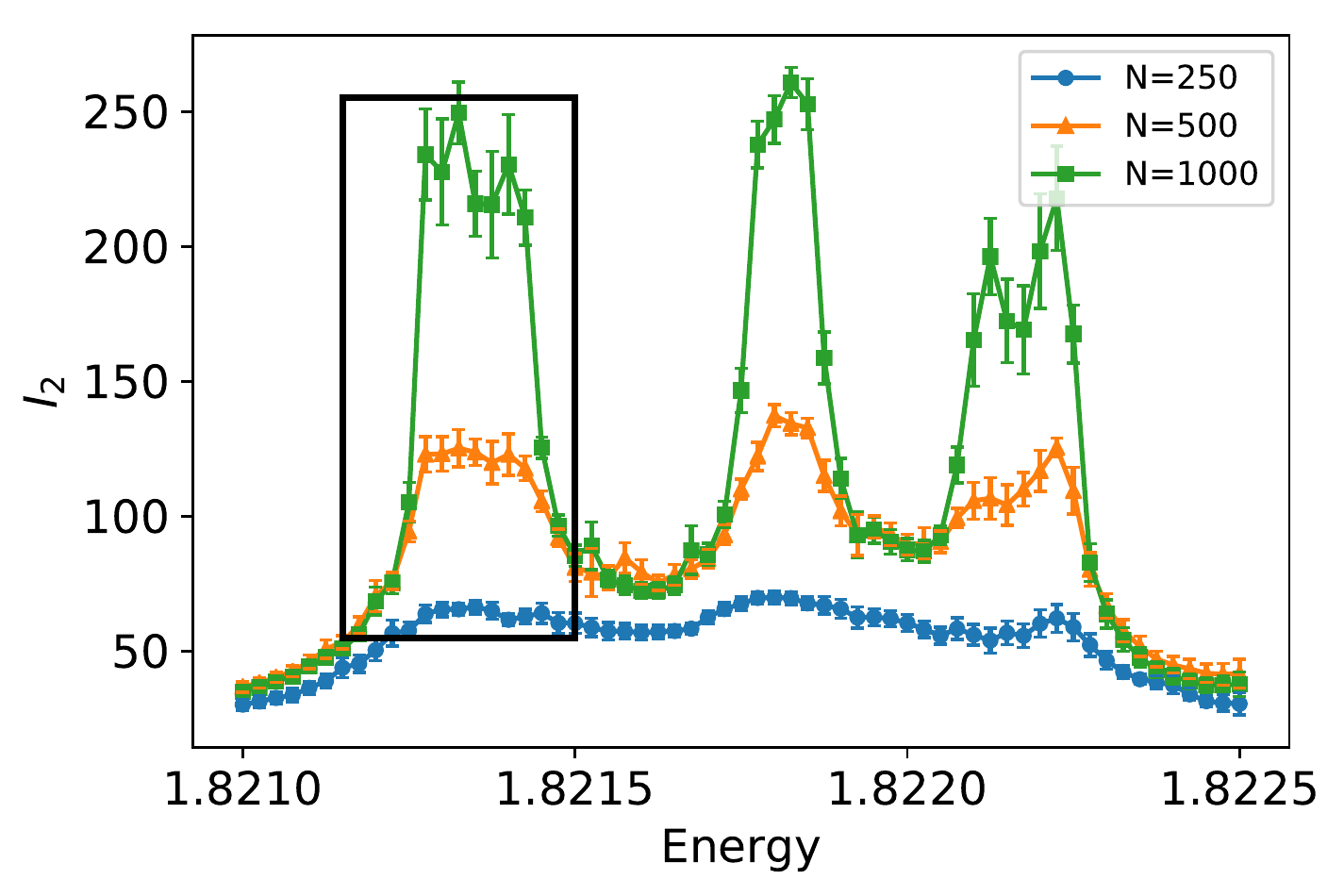}
    \par\end{centering}}
    \subfloat[]{
    \begin{centering}
        \includegraphics[width=0.24\textwidth,height=3cm]{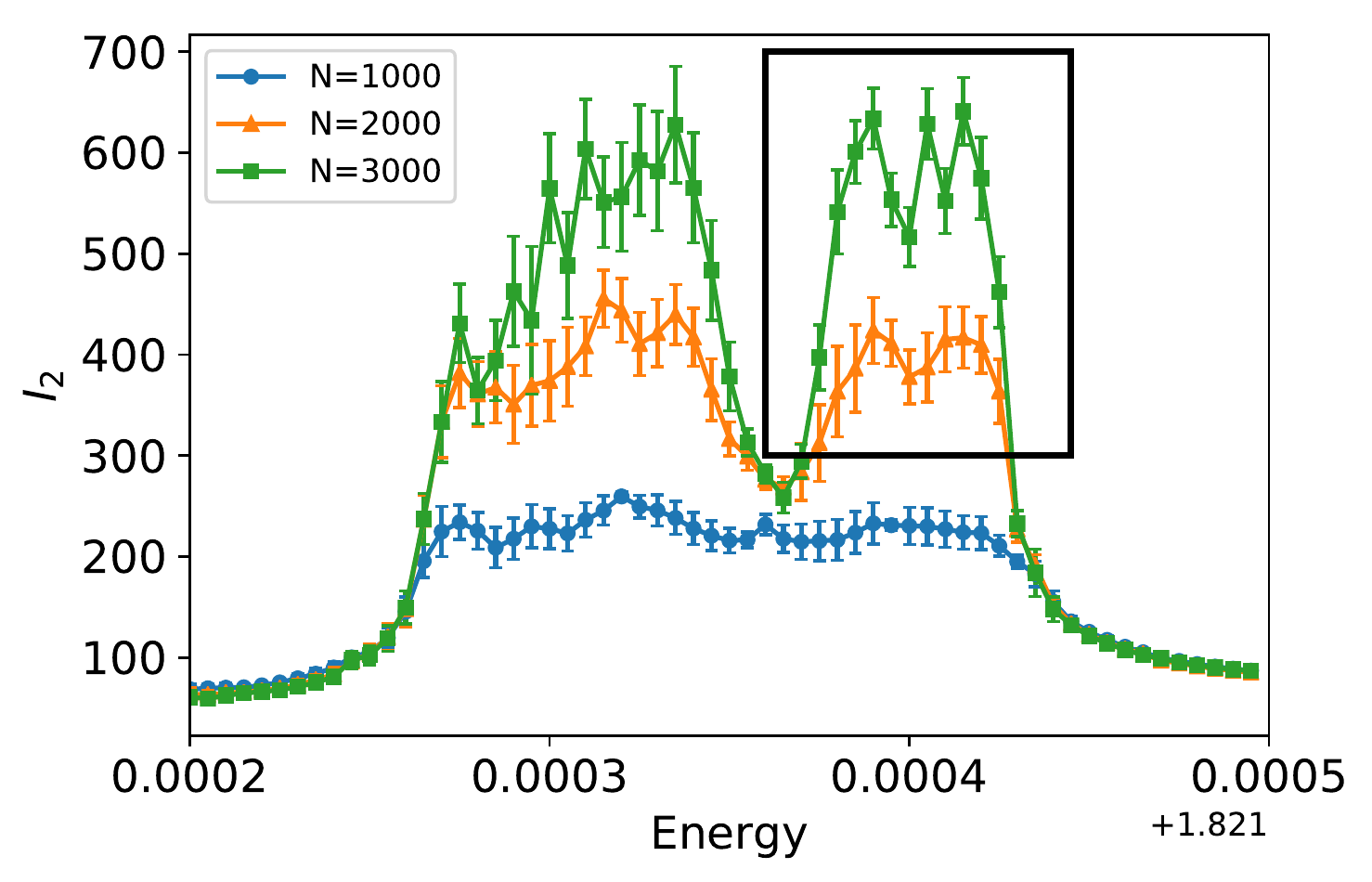}
    \par\end{centering}}
    \subfloat[]{
    \begin{centering}
        \includegraphics[width=0.24\textwidth,height=3cm]{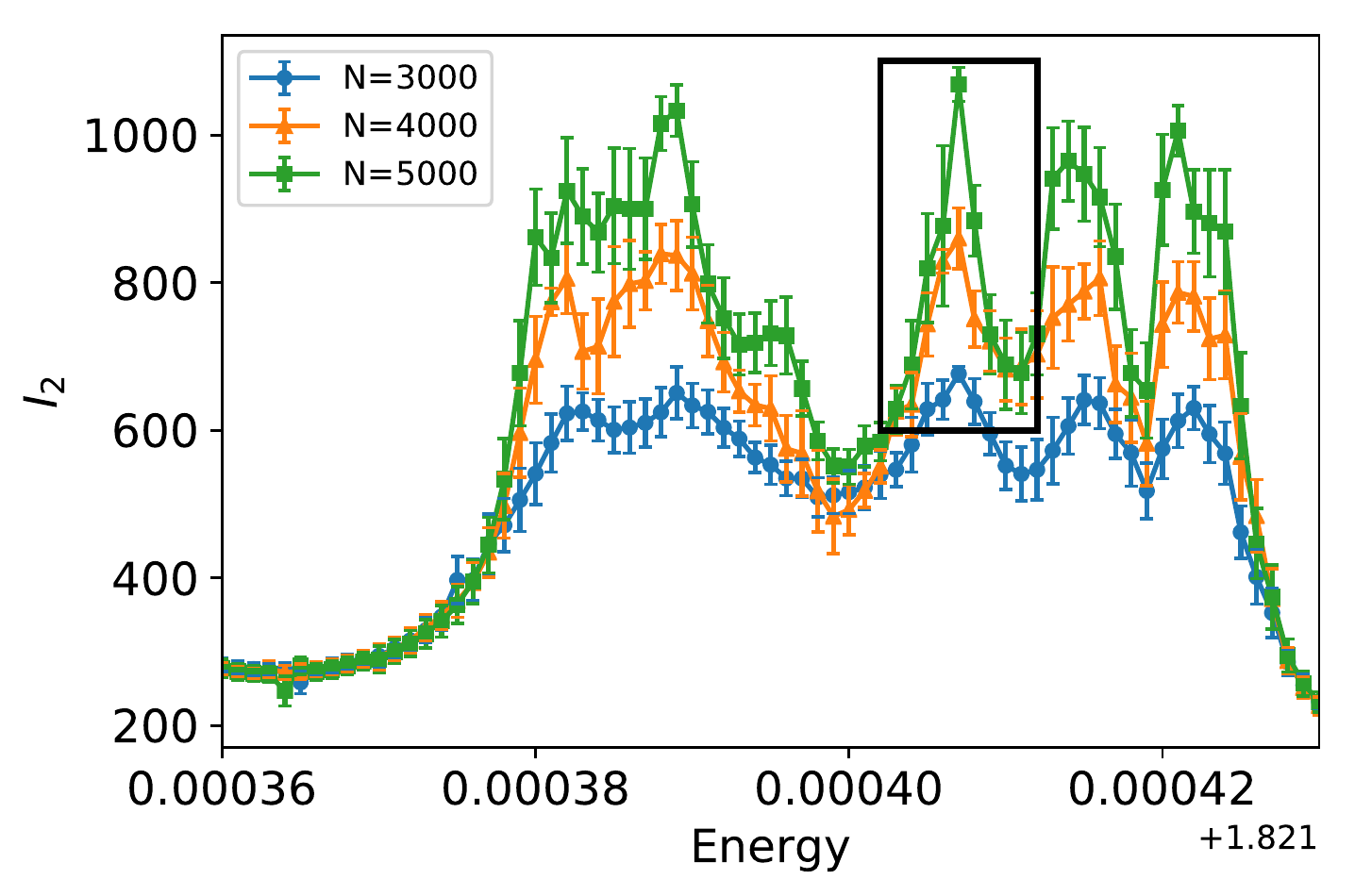}
    \par\end{centering}}
      \subfloat[]{
    \begin{centering}
        \includegraphics[width=0.24\textwidth,height=3cm]{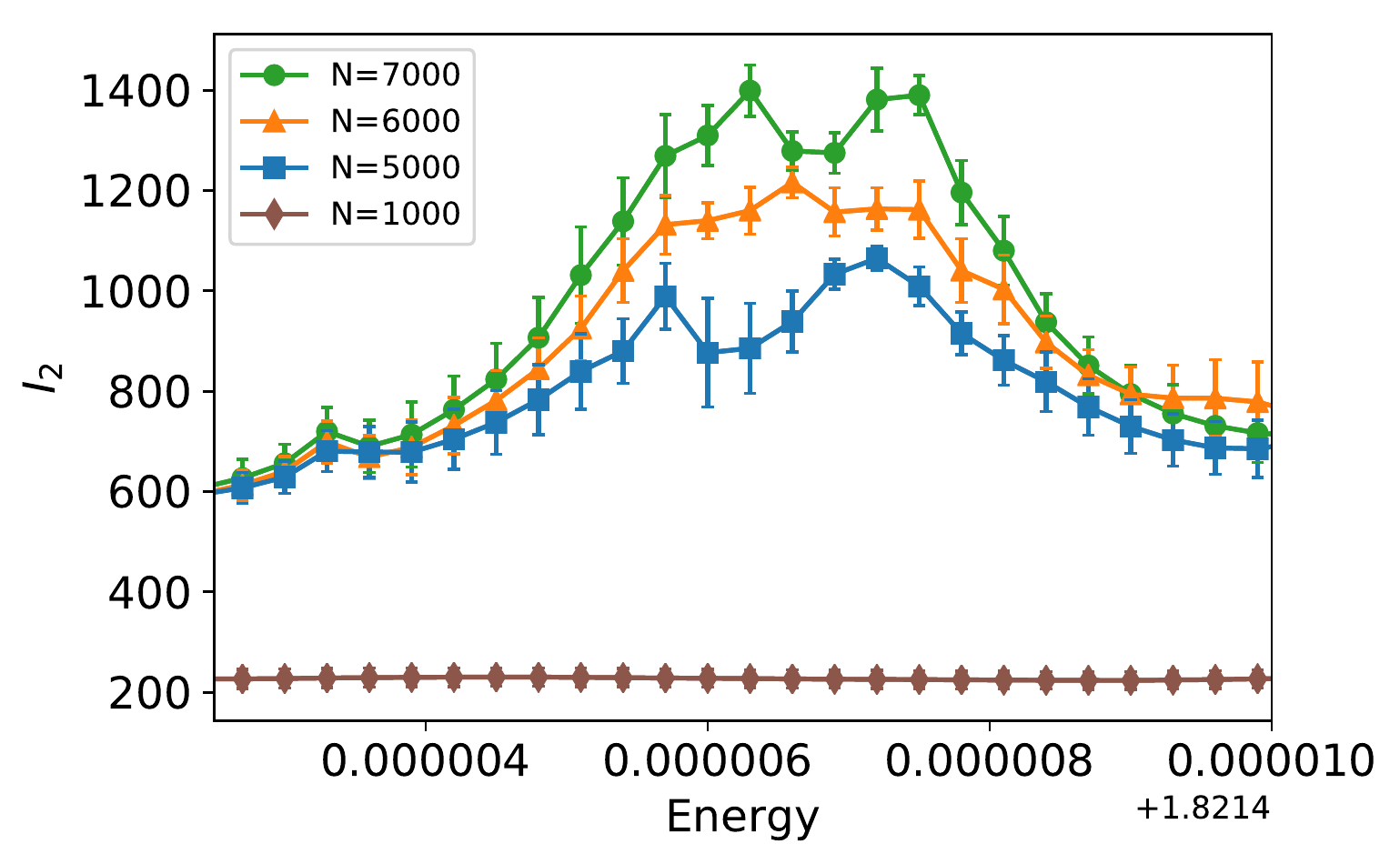}
    \par\end{centering}}
    \caption{(Colour online) Average Green's function participation number $I_2$ around energy $E_1\approx 1.8$ at $u=7.9$. The energy range is zoomed in from left to right, with the maximum system size increasing from $N=1000$ (left) to $N=7000$ (right) and the resolution in energy reaching $\Delta E = 3*10^{-7}$ for the rightmost plot. The errorbars correspond to the disorder average. The peaks of $I_2$ resolve into fine structure with subpeaks upon every iteration of zooming in.}
    \label{fig:5}
\end{figure*}

\begin{figure*}[tbh]
    \centering{}\subfloat[]{
    \begin{centering}
        \includegraphics[width=0.24\textwidth,height=3cm]{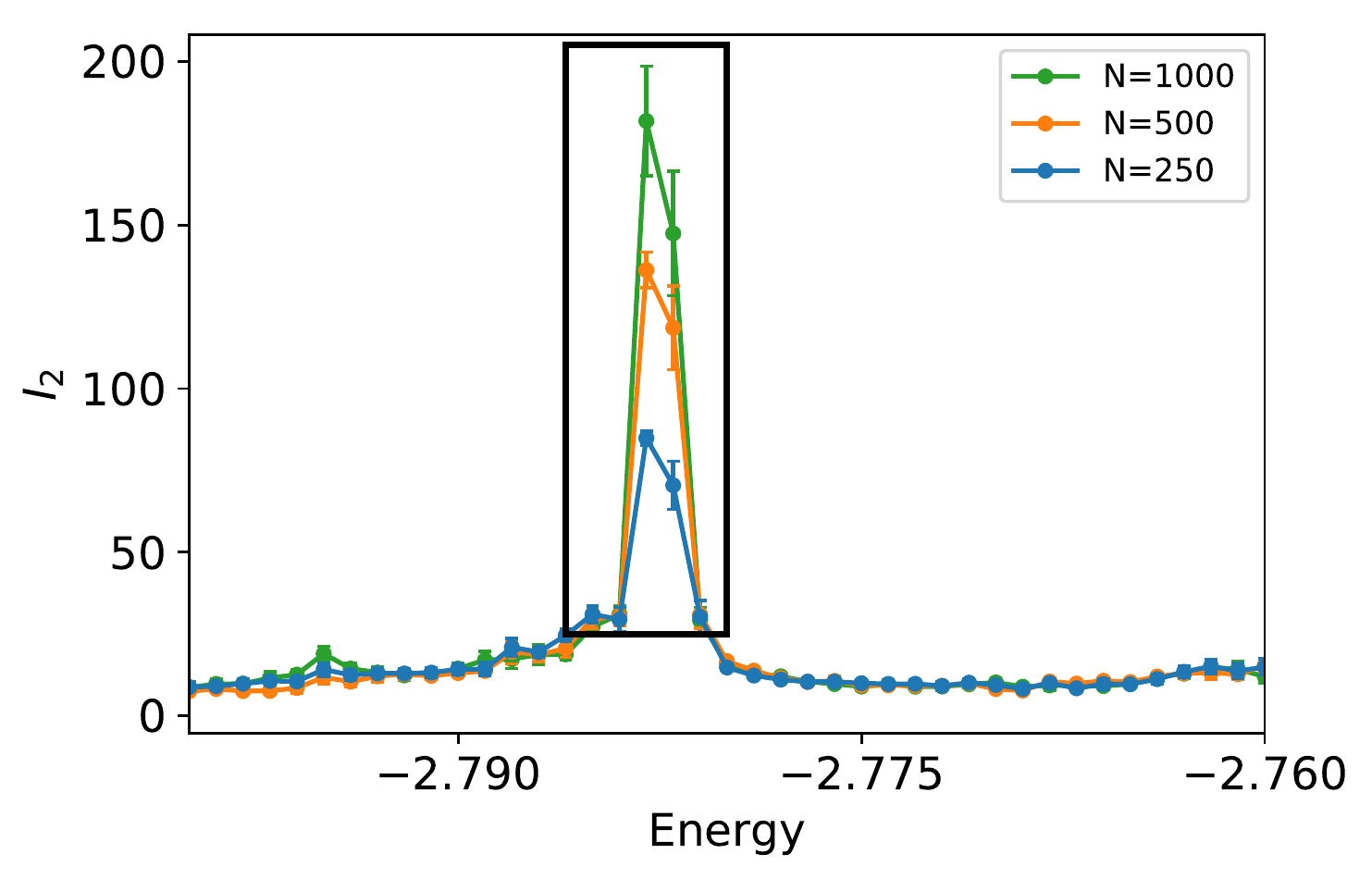}
    \par\end{centering}}
    \subfloat[]{
    \begin{centering}
        \includegraphics[width=0.24\textwidth,height=3cm]{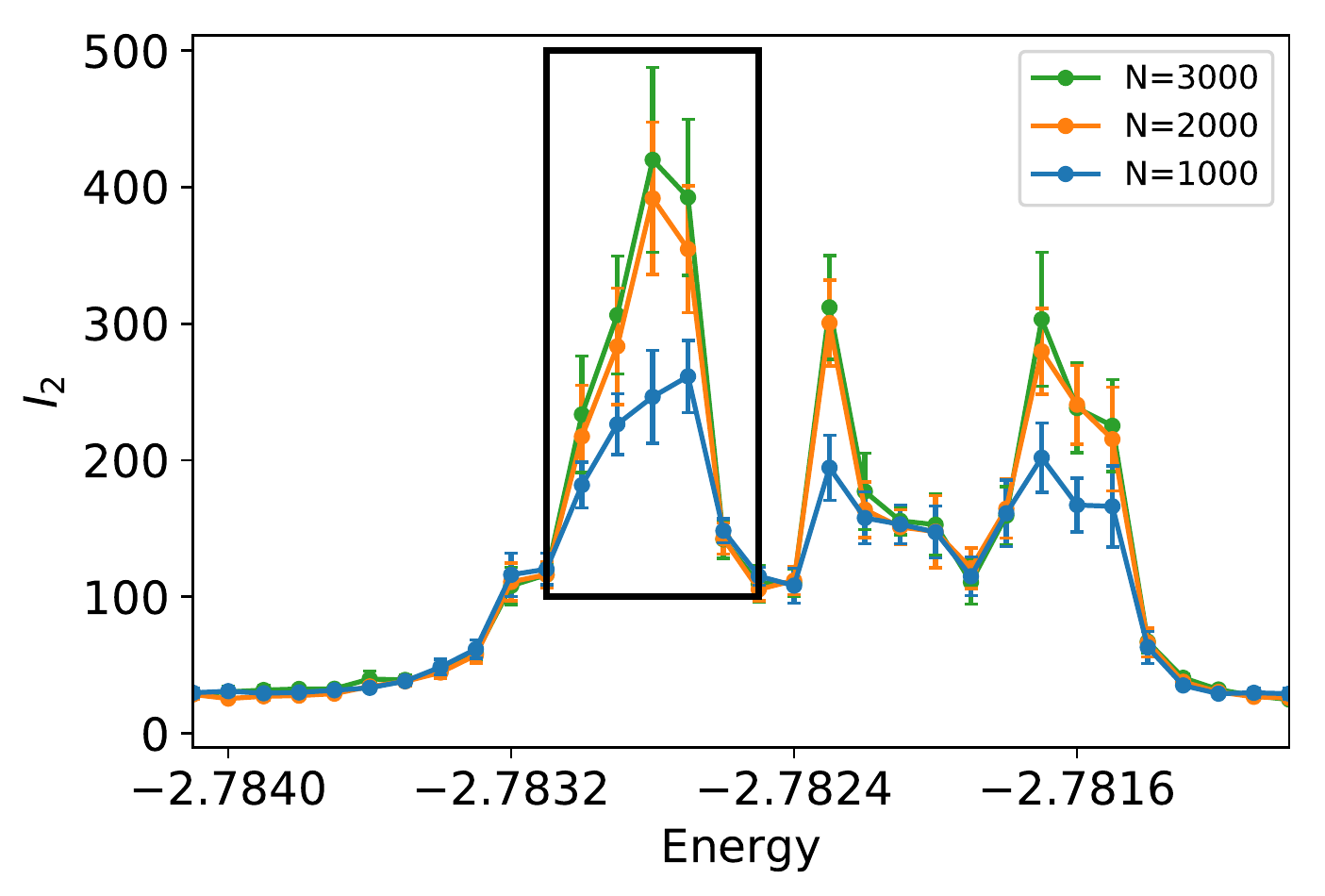}
    \par\end{centering}}
    \subfloat[]{
    \begin{centering}
        \includegraphics[width=0.24\textwidth,height=3cm]{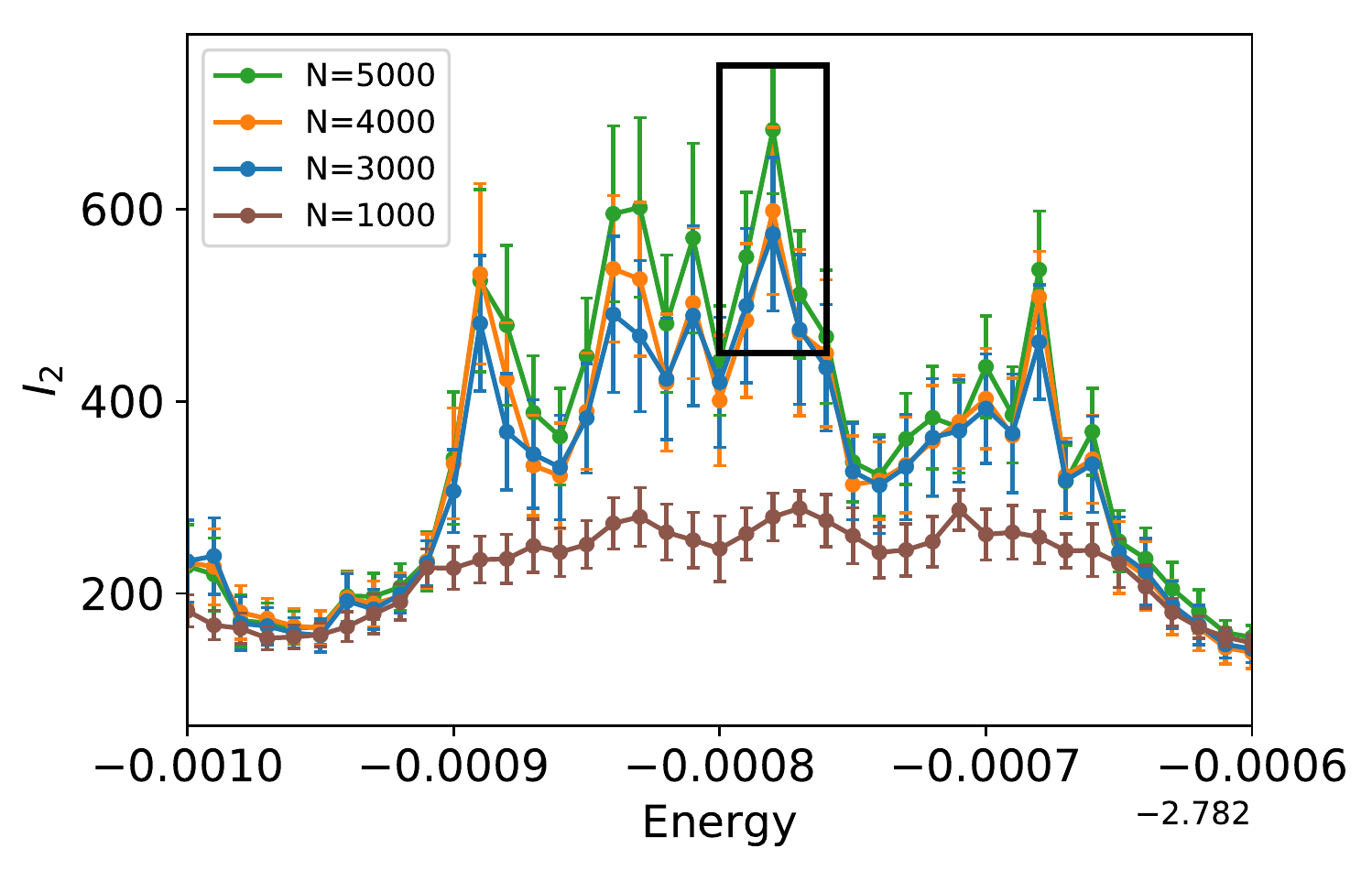}
    \par\end{centering}}
      \subfloat[]{
    \begin{centering}
        \includegraphics[width=0.24\textwidth,height=3cm]{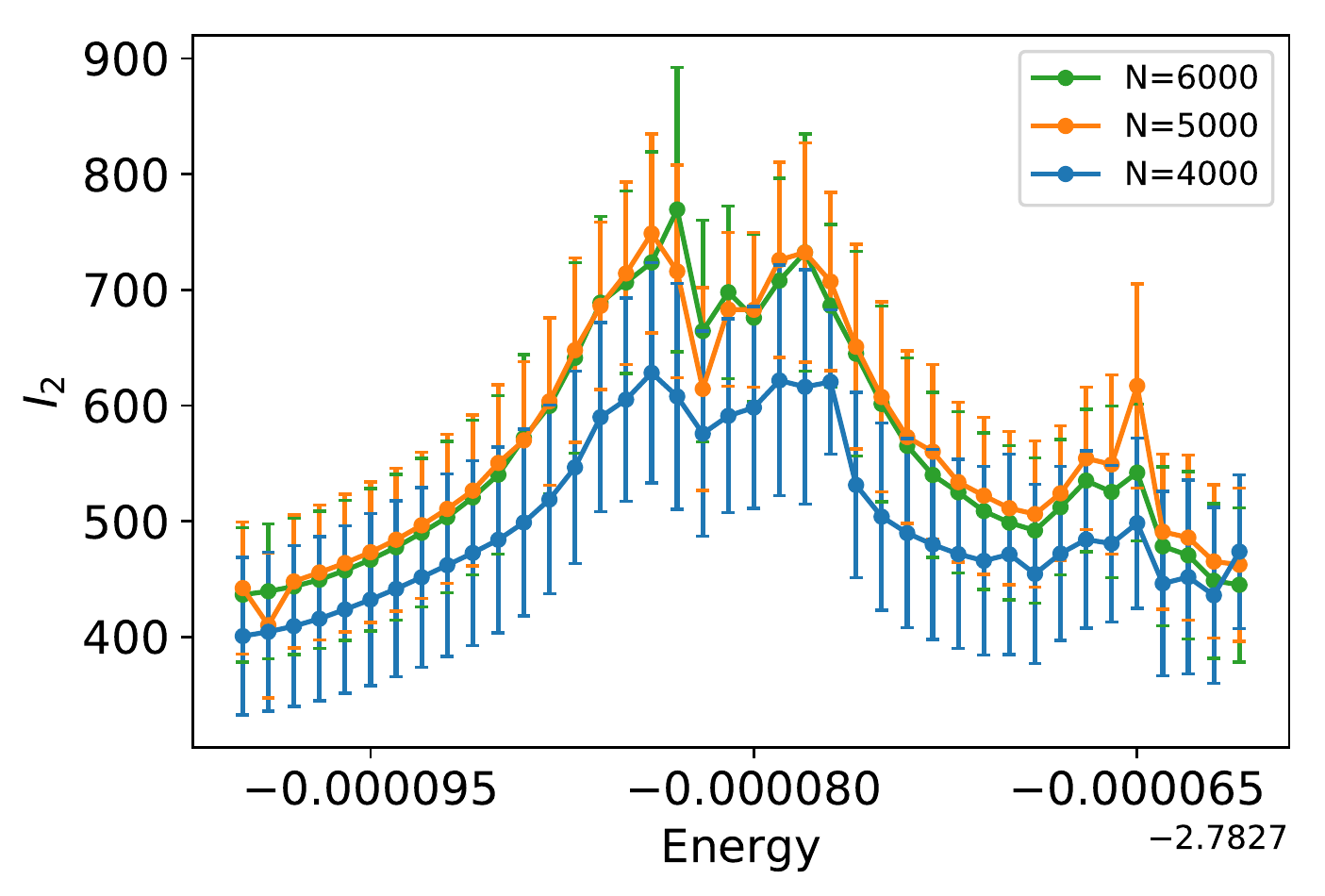}
    \par\end{centering}}
    \caption{(Colour online) Average Green's function participation number $I_2$ around energy $E_2\approx -2.8$ at $u=7.9$. The energy range is zoomed in from left to right, with the maximum system size increasing from $N=1000$ (left) to $N=6000$ (right) and the resolution in energy reaching $\Delta E = 10^{-6}$ for the rightmost plot. The errorbars correspond to the disorder average. The peaks of $I_2$ resolve into fine structure with subpeaks upon every iteration of zooming in.}
    \label{fig:6}
\end{figure*}

To get a better insight into the nature of these emerging states we study the fine structure in the vicinity of the $I_2$ peaks. To extract this fine structure we start with a small system size and identify the peaks of $I_2$ by discretising the energy range. Next we zoom into the energy range around one of the peaks by using a finer energy discretisation. This procedure is repeated several times for increasing system sizes $N$. Such analysis of fine details of the structure of $I_2$ is possible thanks to the usage of the projected Green functions. To be specific for the peak of $I_2$ at $E_1\approx 1.8$, we started with a range or energies $[1.821 :1.8225]$ for the smallest system size $N=250$. We observe the emergence of new peaks which become prominent as the size is increased to $N=500$ and $N=1000$ (Fig.~\ref{fig:5}(a)). Zooming in the energy range around one peak ($E\in [1.8212, 1.8215]$, marked by black rectangular box on Fig.~\ref{fig:5}) the original peak resolves into several peaks for larger system size $N=3000$, Fig.~\ref{fig:5}(b). Repeating this procedure two more times for the peaks marked by the black boxes, we obtain Fig.~\ref{fig:5}(c-d) for $N_\text{max}=7000$. The largest $I_2$ is observed at $E_1=1.8214063$ for $N=7000$. Upon every iteration we observe the emergence of finer structure in $I_2$ as we are zooming in energy. This strongly suggests the fractal nature of participation number $I_2$ as a function of energy $E$ and consequently the spectrum of the delocalised states at these energies.

\begin{figure}[htbh]
    \centering
    \includegraphics[width=0.45\textwidth,angle=0]{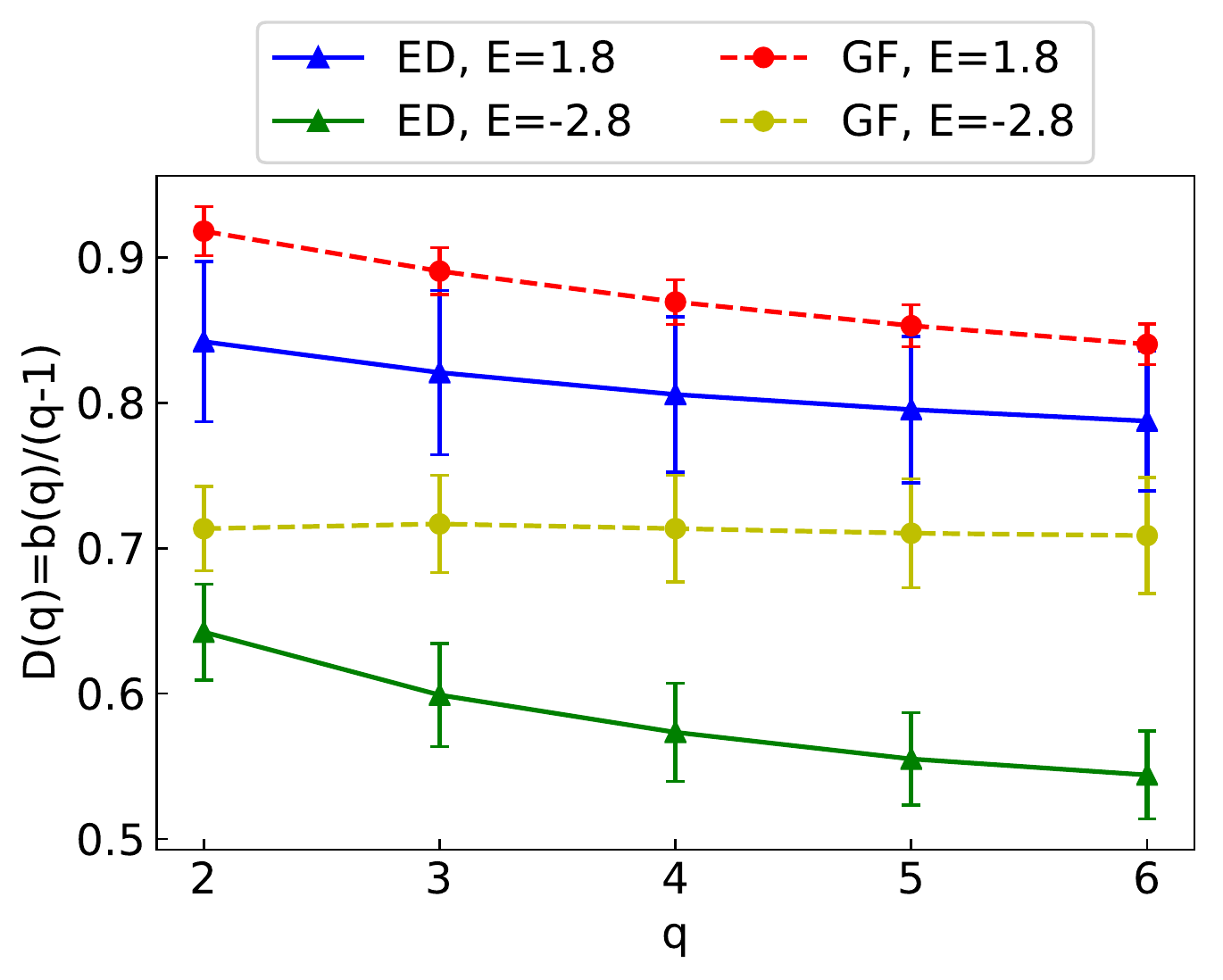}
    \caption{(Colour online) Fractal dimensions $D_q$ (extracted from the Green function participation number $I_q$) and $\mD_q$ (extracted from the participation number $PN_q$) vs $q$ at energies $E_1\approx 1.8$ and $E_2\approx -2.8$. For $E_1$ both methods predict multifractality, while for $E_2$ the projected Green function method underestimates the fractality of the eigenstate.}
    \label{fig:7}
\end{figure}

In the original work, Ref.~\onlinecite{flach2012correlated}, these states were assumed delocalised based on the analysis of wave packet spreading. Subsequent work in Ref.~\onlinecite{frahm2015freed} performed a more detailed analysis and confirmed this conclusion and also provided some indications of fractality of these states based on the fitting i) inverse participation ratio in position representation denoted as $\xi_x$, ii) inverse participation ratio in energy representations, $\xi_E$ (for details see Ref.~\onlinecite{frahm2015freed}). To clarify the fractal nature of these states we consider the largest $I_2$ at energy $E=E_1$ and compute $I_q(N)$ for $q=2,3,4,5,6$ and several system sizes $N$ at this energy. Assuming the multifractal ansatz for the participation number $I_q(N)\sim a N^{D_q(q-1)}$ we extract the fractal dimension $D_q$ from numerical values $I_q(N)$, similarly to how it was done in the single particle case, see Fig.~\ref{fig:2}. The extracted values of $D_q$ are shown as red points (circles) in Fig.~\ref{fig:7} with the error bars of the fit. We observe that $D_q < 1$ and $q$-dependent suggesting that the corresponding eigenstates at this energy are multifractal. In Ref.~\onlinecite{frahm2015freed} a power-law fit of $\xi_x$ and $\xi_E$ with system size $N$ with $N_\text{max}\approx10000$ was computed. The extracted values of the power-law exponents $a_{x,E}<1$ for energies $E=-2.787,1.817$ and interaction $u=7.9$ suggested that these states were fractal.

In the same way, energies around $E_2\approx -2.8$ were analysed, up to system size $N_\text{max}=6000$. The results were averaged over $10$ disorder samples, e.g. values of $\beta$ (see Eq.~\eqref{eq:ham-1p}). The results are shown in Fig.~\ref{fig:6}. We observe larger fluctuations in participation number $I_2$ as compared to $E_1\approx 1.8$ which are shown with the error bars. Also the dependence of $I_2$ on system size $N$ is less prominent as compared to the global maximum of $I_2$ located at $E_1\approx 1.8$ when the energy is zoomed in, even for the largest system size considered (Fig.~\ref{fig:6}c-d). The fractal dimension $D_q$ extracted from $I_q$ shows an almost flat dependence on $q$ (green circles in Fig.~\ref{fig:7}), suggesting only fractal but not multifractal character of the state at this energy.

\begin{figure}[htb]
    \centering
    \includegraphics[width=0.45\textwidth,angle=0]{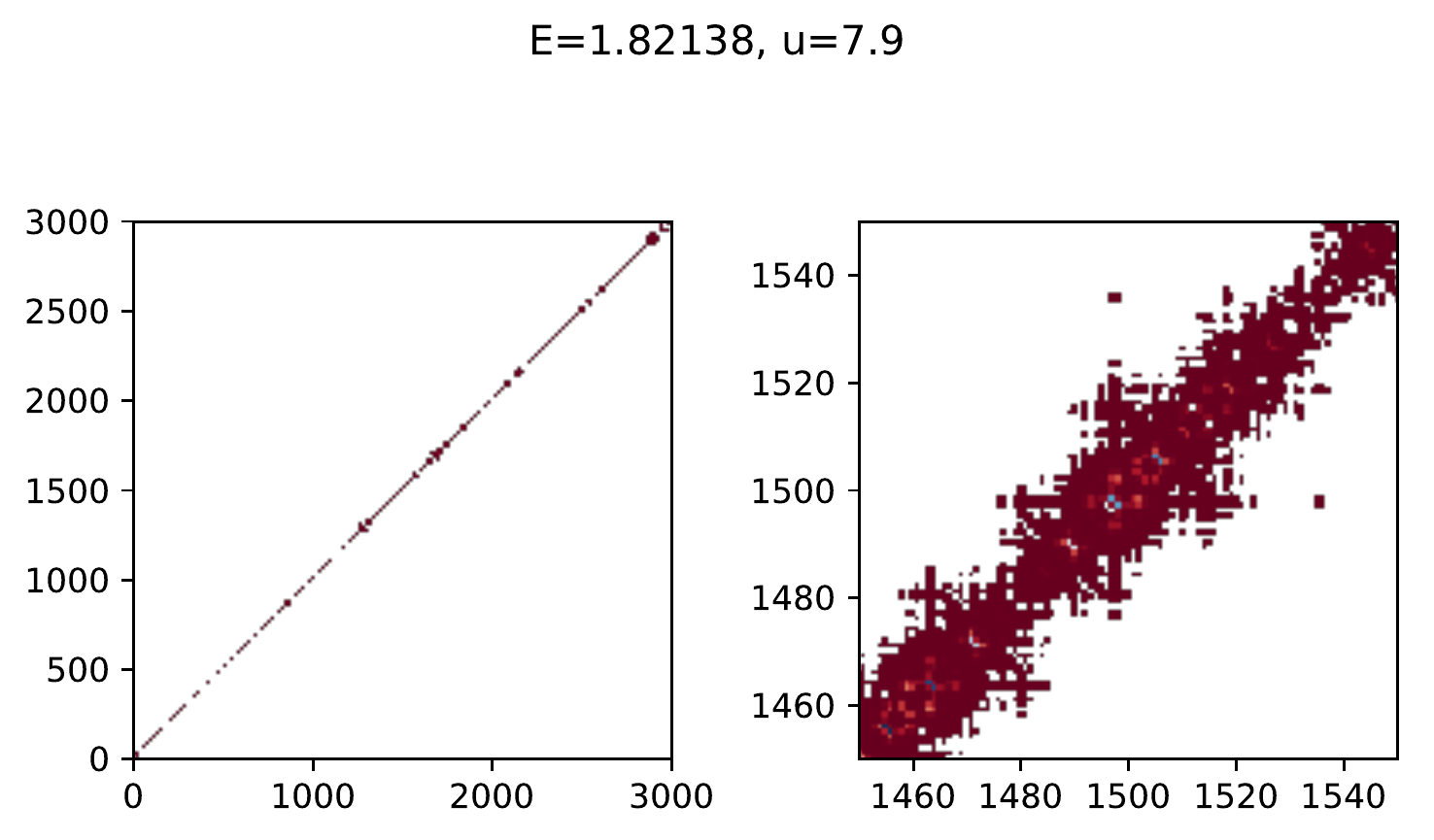}
    \includegraphics[width=0.45\textwidth,angle=0]{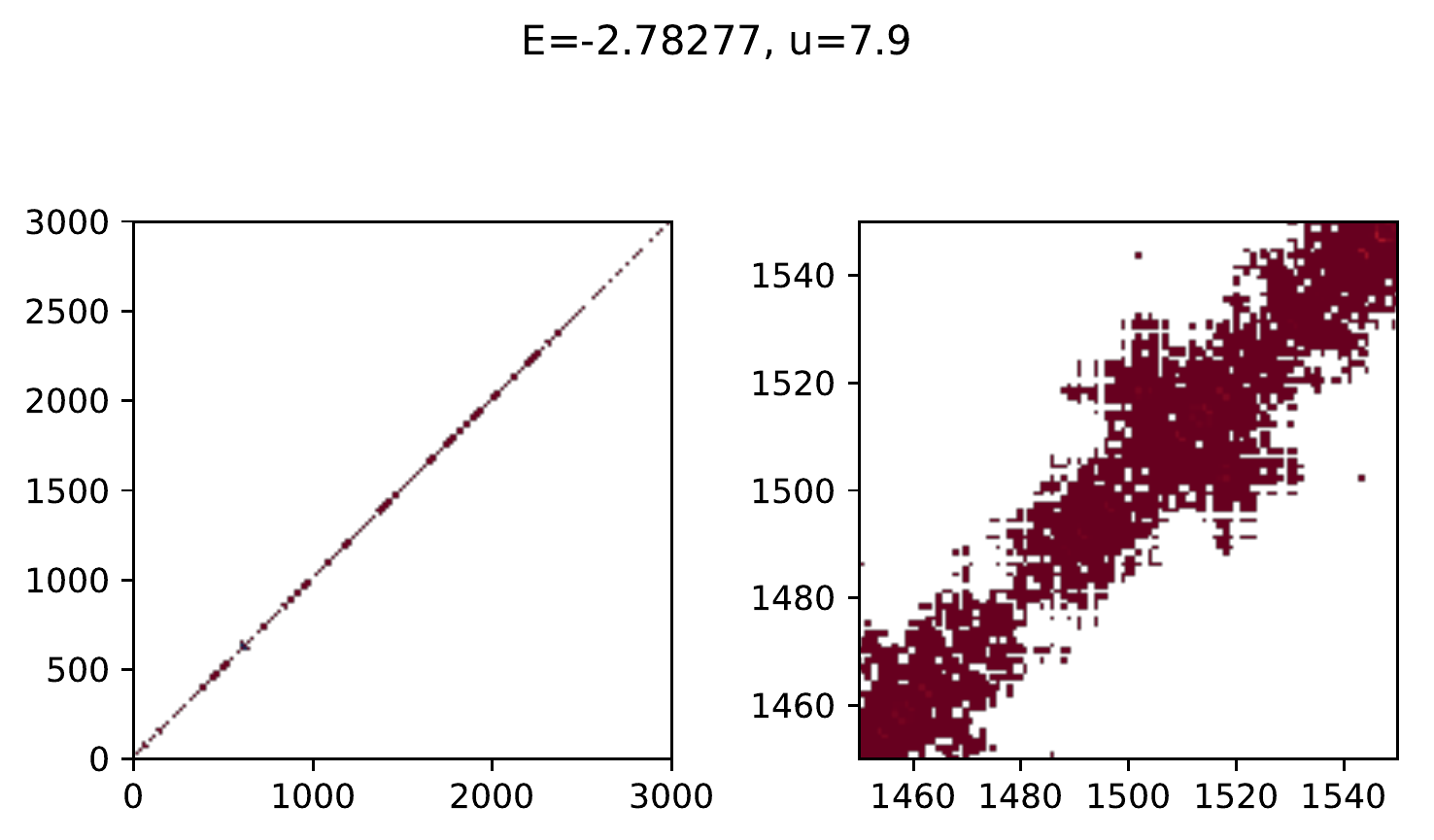}
    \caption{(Colour online) The amplitudes $\vert\Psi(x_1, x_2)\vert$ of eigenstates computed for $N=3000$ at $u=7.9$ and corresponding to the local maxima of $\tpn_2$. The $X$-axis and $Y$-axis denote the positions of the two particles - $x_1$ and $x_2$ - respectively. The larger amplitudes correspond to brighter colour. Values smaller than $10^{-8}$ were discarded. The energies are $E_1\approx 1.8$ (top) and $E_2=-2.78277$ (bottom). Left column: The eigenstate is localised along the main diagonal, e.g. the two particles stick together, but the pattern of the amplitudes along the diagonal is multifractal. Right column: the zoom into the left figure, highlighting the complex, multifractal pattern of the eigenfunction along the diagonal.}
    \label{fig:8}
\end{figure}

The Green function participation number results are indirect, since they do not probe the eigenstates directly. Their advantage is the much lower computational cost for larger system sizes as compared to the exact diagonalisation. Therefore to check our predictions on the fractality of the eigenstates independently we performed sparse diagonalisation around energies $E_1 = 1.8214063$ and $E_2 = -2.782783$, corresponding to the local maxima of $I_2$ for $N_\text{max}=7000$ and $N_\text{max}=6000$ respectively. Among the eigenstates extracted around these two energies, we systematically picked the ones with the largest $PN_2$ for all system sizes $N$ since we aimed at the most delocalised eigenstates embedded into the predominantly localised ones. The power law fits of the participation number moments, $\text{PN}_q(N)\propto N^{\mD_q(q-1)}$ were calculated. The resulting values of $\mD_q$ are shown in Fig.~\ref{fig:7} as blue ($E=1.8$) and green ($E=-2.8$) solid lines with triangular points. The dashed lines with points show $D_q$ evaluated from the Green function participation numbers, red for $E_1=1.8$ and yellow for $E_2=-2.8$. We see that although the values of $D_q$ and $\mD_q$ do not always agree perfectly, nevertheless $D_q$ and $\mD_q$ imply at least fractality of the eigenstates that were previously considered delocalised.~\cite{flach2012correlated} This also provides yet another evidence for the validity of $I_q$ as a measure of localisation of eigenstates.

We elaborate further on the character of these fractal states appearing around $E_{1,2}$. Since the appearance of these states relies crucially on the interaction, we expect them to have a peculiar spatial pattern of the wavefunction amplitudes. Indeed we can construct many approximate localised eigenstates with two particles separated by one or more localisation lengths $\xi_1$. Therefore the fractal states should have the two particles separated by at most the single particles localisation length $\xi_1$. If we visualise the amplitudes of the two particles eigenfuncion $\vert\Psi(x_1, x_2)\vert$ on a square lattice with coordinates $x_1$, $x_2$, that correspond to the positions of the two particles, we expect the fractal states to be localised along the main diagonal $x_1=x_2$, the fractal structure translating into some complicated pattern along the main diagonal. To verify this hypothesis we plotted two exact eigenstates with the largest $PN_2$ for $N=3000$ in Fig.~\ref{fig:8} ($u=7.9$ and $E_1\approx 1.8$ (top) and $E_2\approx -2.78$). The axes denote the position of each of the two particles. We truncated amplitudes $\vert\Psi(x_1, x_2)\vert < 10^{-8}$ on the plots. These plots fully confirm our hypothesis outlined above, with most weight concentrated along the main diagonal, i.e. both particles being close to each other.

\section{Conclusions}

To conclude, we have shown that previously discovered metallic states of two interacting particles in an AA chain in the insulating single-particle region have a fractal structure. Furthermore unlike previous claims we find that these states are multifractal. This is verified by computing participation numbers from projected GF as well as from exact diagonalisation. An interesting open problem is the fate of these multifractal states at finite density where many-body localisation was reported at half-filling.~\cite{iyer2013many} 

As a side effect, we demonstrated that the projected Green functions can be used as a first probe to check the nature of eigenstates in an interacting Hamiltonian system having the advantage that larger system sizes can be targeted as compared to the computationally challenging exact diagonalisation.

\begin{acknowledgments}
        This work was supported by the Institute for Basic Science in Korea (IBS-R024-D1). 
\end{acknowledgments}

\bibliography{general,mbl}

\begin{thebibliography}{19}%
\makeatletter
\providecommand \@ifxundefined [1]{%
 \@ifx{#1\undefined}
}%
\providecommand \@ifnum [1]{%
 \ifnum #1\expandafter \@firstoftwo
 \else \expandafter \@secondoftwo
 \fi
}%
\providecommand \@ifx [1]{%
 \ifx #1\expandafter \@firstoftwo
 \else \expandafter \@secondoftwo
 \fi
}%
\providecommand \natexlab [1]{#1}%
\providecommand \enquote  [1]{``#1''}%
\providecommand \bibnamefont  [1]{#1}%
\providecommand \bibfnamefont [1]{#1}%
\providecommand \citenamefont [1]{#1}%
\providecommand \href@noop [0]{\@secondoftwo}%
\providecommand \href [0]{\begingroup \@sanitize@url \@href}%
\providecommand \@href[1]{\@@startlink{#1}\@@href}%
\providecommand \@@href[1]{\endgroup#1\@@endlink}%
\providecommand \@sanitize@url [0]{\catcode `\\12\catcode `\$12\catcode
  `\&12\catcode `\#12\catcode `\^12\catcode `\_12\catcode `\%12\relax}%
\providecommand \@@startlink[1]{}%
\providecommand \@@endlink[0]{}%
\providecommand \url  [0]{\begingroup\@sanitize@url \@url }%
\providecommand \@url [1]{\endgroup\@href {#1}{\urlprefix }}%
\providecommand \urlprefix  [0]{URL }%
\providecommand \Eprint [0]{\href }%
\providecommand \doibase [0]{http://dx.doi.org/}%
\providecommand \selectlanguage [0]{\@gobble}%
\providecommand \bibinfo  [0]{\@secondoftwo}%
\providecommand \bibfield  [0]{\@secondoftwo}%
\providecommand \translation [1]{[#1]}%
\providecommand \BibitemOpen [0]{}%
\providecommand \bibitemStop [0]{}%
\providecommand \bibitemNoStop [0]{.\EOS\space}%
\providecommand \EOS [0]{\spacefactor3000\relax}%
\providecommand \BibitemShut  [1]{\csname bibitem#1\endcsname}%
\let\auto@bib@innerbib\@empty
\bibitem [{\citenamefont {Anderson}(1958)}]{anderson1958absence}%
  \BibitemOpen
  \bibfield  {author} {\bibinfo {author} {\bibfnamefont {P.~W.}\ \bibnamefont
  {Anderson}},\ }\bibfield  {title} {\enquote {\bibinfo {title} {Absence of
  diffusion in certain random lattices},}\ }\href {\doibase
  10.1103/PhysRev.109.1492} {\bibfield  {journal} {\bibinfo  {journal} {Phys.
  Rev.}\ }\textbf {\bibinfo {volume} {109}},\ \bibinfo {pages} {1492--1505}
  (\bibinfo {year} {1958})}\BibitemShut {NoStop}%
\bibitem [{\citenamefont {Kramer}\ and\ \citenamefont
  {MacKinnon}(1993)}]{kramer1993localization}%
  \BibitemOpen
  \bibfield  {author} {\bibinfo {author} {\bibfnamefont {B.}~\bibnamefont
  {Kramer}}\ and\ \bibinfo {author} {\bibfnamefont {A.}~\bibnamefont
  {MacKinnon}},\ }\bibfield  {title} {\enquote {\bibinfo {title} {Localization:
  theory and experiment},}\ }\href {\doibase 10.1088/0034-4885/56/12/001}
  {\bibfield  {journal} {\bibinfo  {journal} {Rep. Prog. Phys.}\ }\textbf
  {\bibinfo {volume} {56}},\ \bibinfo {pages} {1469--1564} (\bibinfo {year}
  {1993})}\BibitemShut {NoStop}%
\bibitem [{\citenamefont {Basko}\ \emph {et~al.}(2006)\citenamefont {Basko},
  \citenamefont {Aleiner},\ and\ \citenamefont {Altshuler}}]{basko2006metal}%
  \BibitemOpen
  \bibfield  {author} {\bibinfo {author} {\bibfnamefont {D.M.}\ \bibnamefont
  {Basko}}, \bibinfo {author} {\bibfnamefont {I.L.}\ \bibnamefont {Aleiner}}, \
  and\ \bibinfo {author} {\bibfnamefont {B.L.}\ \bibnamefont {Altshuler}},\
  }\bibfield  {title} {\enquote {\bibinfo {title} {Metal–insulator transition
  in a weakly interacting many-electron system with localized single-particle
  states},}\ }\href {\doibase 10.1016/j.aop.2005.11.014} {\bibfield  {journal}
  {\bibinfo  {journal} {Ann. Phys.}\ }\textbf {\bibinfo {volume} {321}},\
  \bibinfo {pages} {1126 -- 1205} (\bibinfo {year} {2006})}\BibitemShut
  {NoStop}%
\bibitem [{\citenamefont {Abanin}\ \emph {et~al.}(2019)\citenamefont {Abanin},
  \citenamefont {Altman}, \citenamefont {Bloch},\ and\ \citenamefont
  {Serbyn}}]{abanin2019colloquium}%
  \BibitemOpen
  \bibfield  {author} {\bibinfo {author} {\bibfnamefont {Dmitry~A.}\
  \bibnamefont {Abanin}}, \bibinfo {author} {\bibfnamefont {Ehud}\ \bibnamefont
  {Altman}}, \bibinfo {author} {\bibfnamefont {Immanuel}\ \bibnamefont
  {Bloch}}, \ and\ \bibinfo {author} {\bibfnamefont {Maksym}\ \bibnamefont
  {Serbyn}},\ }\bibfield  {title} {\enquote {\bibinfo {title} {Colloquium:
  Many-body localization, thermalization, and entanglement},}\ }\href {\doibase
  10.1103/RevModPhys.91.021001} {\bibfield  {journal} {\bibinfo  {journal}
  {Rev. Mod. Phys.}\ }\textbf {\bibinfo {volume} {91}},\ \bibinfo {pages}
  {021001} (\bibinfo {year} {2019})}\BibitemShut {NoStop}%
\bibitem [{\citenamefont {Alet}\ and\ \citenamefont
  {Laflorencie}(2018)}]{fabien2018many}%
  \BibitemOpen
  \bibfield  {author} {\bibinfo {author} {\bibfnamefont {Fabien}\ \bibnamefont
  {Alet}}\ and\ \bibinfo {author} {\bibfnamefont {Nicolas}\ \bibnamefont
  {Laflorencie}},\ }\bibfield  {title} {\enquote {\bibinfo {title} {Many-body
  localization: An introduction and selected topics},}\ }\href {\doibase
  https://doi.org/10.1016/j.crhy.2018.03.003} {\bibfield  {journal} {\bibinfo
  {journal} {Comptes Rend. Phys.}\ }\textbf {\bibinfo {volume} {19}},\ \bibinfo
  {pages} {498 -- 525} (\bibinfo {year} {2018})}\BibitemShut {NoStop}%
\bibitem [{\citenamefont {Fukuyama}\ \emph {et~al.}(1973)\citenamefont
  {Fukuyama}, \citenamefont {Bari},\ and\ \citenamefont
  {Fogedby}}]{fukuyama1973tightly}%
  \BibitemOpen
  \bibfield  {author} {\bibinfo {author} {\bibfnamefont {Hidetoshi}\
  \bibnamefont {Fukuyama}}, \bibinfo {author} {\bibfnamefont {Robert~A.}\
  \bibnamefont {Bari}}, \ and\ \bibinfo {author} {\bibfnamefont {Hans~C.}\
  \bibnamefont {Fogedby}},\ }\bibfield  {title} {\enquote {\bibinfo {title}
  {Tightly bound electrons in a uniform electric field},}\ }\href {\doibase
  10.1103/PhysRevB.8.5579} {\bibfield  {journal} {\bibinfo  {journal} {Phys.
  Rev. B}\ }\textbf {\bibinfo {volume} {8}},\ \bibinfo {pages} {5579--5586}
  (\bibinfo {year} {1973})}\BibitemShut {NoStop}%
\bibitem [{\citenamefont {Aubry}\ and\ \citenamefont
  {Andr{\'e}}(1980)}]{aubry1980analyticity}%
  \BibitemOpen
  \bibfield  {author} {\bibinfo {author} {\bibfnamefont {Serge}\ \bibnamefont
  {Aubry}}\ and\ \bibinfo {author} {\bibfnamefont {Gilles}\ \bibnamefont
  {Andr{\'e}}},\ }\bibfield  {title} {\enquote {\bibinfo {title} {Analyticity
  breaking and anderson localization in incommensurate lattices},}\ }\href@noop
  {} {\bibfield  {journal} {\bibinfo  {journal} {Ann. Israel Phys. Soc}\
  }\textbf {\bibinfo {volume} {3}},\ \bibinfo {pages} {18} (\bibinfo {year}
  {1980})}\BibitemShut {NoStop}%
\bibitem [{\citenamefont {Shepelyansky}(1994)}]{shepelyansky1994coherent}%
  \BibitemOpen
  \bibfield  {author} {\bibinfo {author} {\bibfnamefont {D.~L.}\ \bibnamefont
  {Shepelyansky}},\ }\bibfield  {title} {\enquote {\bibinfo {title} {Coherent
  propagation of two interacting particles in a random potential},}\ }\href
  {\doibase 10.1103/PhysRevLett.73.2607} {\bibfield  {journal} {\bibinfo
  {journal} {Phys. Rev. Lett.}\ }\textbf {\bibinfo {volume} {73}},\ \bibinfo
  {pages} {2607--2610} (\bibinfo {year} {1994})}\BibitemShut {NoStop}%
\bibitem [{\citenamefont {Frahm}(1999)}]{frahm1999interaction}%
  \BibitemOpen
  \bibfield  {author} {\bibinfo {author} {\bibfnamefont {K.M.}\ \bibnamefont
  {Frahm}},\ }\bibfield  {title} {\enquote {\bibinfo {title} {Interaction
  induced delocalization of two particles: large system size calculations and
  dependence on interaction strength},}\ }\href {\doibase
  10.1007/s100510050866} {\bibfield  {journal} {\bibinfo  {journal} {Eur. Phys.
  J. B}\ }\textbf {\bibinfo {volume} {10}},\ \bibinfo {pages} {371--378}
  (\bibinfo {year} {1999})}\BibitemShut {NoStop}%
\bibitem [{\citenamefont {Krimer}\ \emph {et~al.}(2011)\citenamefont {Krimer},
  \citenamefont {Khomeriki},\ and\ \citenamefont {Flach}}]{krimer2011two}%
  \BibitemOpen
  \bibfield  {author} {\bibinfo {author} {\bibfnamefont {D.~O.}\ \bibnamefont
  {Krimer}}, \bibinfo {author} {\bibfnamefont {R.}~\bibnamefont {Khomeriki}}, \
  and\ \bibinfo {author} {\bibfnamefont {S.}~\bibnamefont {Flach}},\ }\bibfield
   {title} {\enquote {\bibinfo {title} {Two interacting particles in a random
  potential},}\ }\href {\doibase 10.1134/S0021364011170097} {\bibfield
  {journal} {\bibinfo  {journal} {JETP Lett.}\ }\textbf {\bibinfo {volume}
  {94}},\ \bibinfo {pages} {406--412} (\bibinfo {year} {2011})}\BibitemShut
  {NoStop}%
\bibitem [{\citenamefont {Thongjaomayum}\ \emph {et~al.}(2019)\citenamefont
  {Thongjaomayum}, \citenamefont {Andreanov}, \citenamefont {Engl},\ and\
  \citenamefont {Flach}}]{thongjaomayum2019taming}%
  \BibitemOpen
  \bibfield  {author} {\bibinfo {author} {\bibfnamefont {Diana}\ \bibnamefont
  {Thongjaomayum}}, \bibinfo {author} {\bibfnamefont {Alexei}\ \bibnamefont
  {Andreanov}}, \bibinfo {author} {\bibfnamefont {Thomas}\ \bibnamefont
  {Engl}}, \ and\ \bibinfo {author} {\bibfnamefont {Sergej}\ \bibnamefont
  {Flach}},\ }\bibfield  {title} {\enquote {\bibinfo {title} {Taming two
  interacting particles with disorder},}\ }\href {\doibase
  10.1103/PhysRevB.100.224203} {\bibfield  {journal} {\bibinfo  {journal}
  {Phys. Rev. B}\ }\textbf {\bibinfo {volume} {100}},\ \bibinfo {pages}
  {224203} (\bibinfo {year} {2019})}\BibitemShut {NoStop}%
\bibitem [{\citenamefont {Khomeriki}\ \emph {et~al.}(2010)\citenamefont
  {Khomeriki}, \citenamefont {Krimer}, \citenamefont {Haque},\ and\
  \citenamefont {Flach}}]{khomeriki2010interaction}%
  \BibitemOpen
  \bibfield  {author} {\bibinfo {author} {\bibfnamefont {Ramaz}\ \bibnamefont
  {Khomeriki}}, \bibinfo {author} {\bibfnamefont {Dmitry~O.}\ \bibnamefont
  {Krimer}}, \bibinfo {author} {\bibfnamefont {Masudul}\ \bibnamefont {Haque}},
  \ and\ \bibinfo {author} {\bibfnamefont {Sergej}\ \bibnamefont {Flach}},\
  }\bibfield  {title} {\enquote {\bibinfo {title} {Interaction-induced
  fractional bloch and tunneling oscillations},}\ }\href {\doibase
  10.1103/PhysRevA.81.065601} {\bibfield  {journal} {\bibinfo  {journal} {Phys.
  Rev. A}\ }\textbf {\bibinfo {volume} {81}},\ \bibinfo {pages} {065601}
  (\bibinfo {year} {2010})}\BibitemShut {NoStop}%
\bibitem [{\citenamefont {Flach}\ \emph {et~al.}(2012)\citenamefont {Flach},
  \citenamefont {Ivanchenko},\ and\ \citenamefont
  {Khomeriki}}]{flach2012correlated}%
  \BibitemOpen
  \bibfield  {author} {\bibinfo {author} {\bibfnamefont {Sergej}\ \bibnamefont
  {Flach}}, \bibinfo {author} {\bibfnamefont {Mikhail}\ \bibnamefont
  {Ivanchenko}}, \ and\ \bibinfo {author} {\bibfnamefont {Ramaz}\ \bibnamefont
  {Khomeriki}},\ }\bibfield  {title} {\enquote {\bibinfo {title} {Correlated
  metallic two-particle bound states in quasiperiodic chains},}\ }\href
  {\doibase 10.1209/0295-5075/98/66002} {\bibfield  {journal} {\bibinfo
  {journal} {EPL}\ }\textbf {\bibinfo {volume} {98}},\ \bibinfo {pages} {66002}
  (\bibinfo {year} {2012})}\BibitemShut {NoStop}%
\bibitem [{\citenamefont {Frahm}\ and\ \citenamefont
  {Shepelyansky}(2015)}]{frahm2015freed}%
  \BibitemOpen
  \bibfield  {author} {\bibinfo {author} {\bibfnamefont {Klaus~M.}\
  \bibnamefont {Frahm}}\ and\ \bibinfo {author} {\bibfnamefont {Dima~L.}\
  \bibnamefont {Shepelyansky}},\ }\bibfield  {title} {\enquote {\bibinfo
  {title} {Freed by interaction kinetic states in the harper model},}\ }\href
  {\doibase 10.1140/epjb/e2015-60733-9} {\bibfield  {journal} {\bibinfo
  {journal} {Eur. Phys. J. B}\ }\textbf {\bibinfo {volume} {88}},\ \bibinfo
  {pages} {337} (\bibinfo {year} {2015})}\BibitemShut {NoStop}%
\bibitem [{\citenamefont {Schreiber}\ \emph {et~al.}(2015)\citenamefont
  {Schreiber}, \citenamefont {Hodgman}, \citenamefont {Bordia}, \citenamefont
  {L{\"u}schen}, \citenamefont {Fischer}, \citenamefont {Vosk}, \citenamefont
  {Altman}, \citenamefont {Schneider},\ and\ \citenamefont
  {Bloch}}]{schreiber2015observation}%
  \BibitemOpen
  \bibfield  {author} {\bibinfo {author} {\bibfnamefont {Michael}\ \bibnamefont
  {Schreiber}}, \bibinfo {author} {\bibfnamefont {Sean~S.}\ \bibnamefont
  {Hodgman}}, \bibinfo {author} {\bibfnamefont {Pranjal}\ \bibnamefont
  {Bordia}}, \bibinfo {author} {\bibfnamefont {Henrik~P.}\ \bibnamefont
  {L{\"u}schen}}, \bibinfo {author} {\bibfnamefont {Mark~H.}\ \bibnamefont
  {Fischer}}, \bibinfo {author} {\bibfnamefont {Ronen}\ \bibnamefont {Vosk}},
  \bibinfo {author} {\bibfnamefont {Ehud}\ \bibnamefont {Altman}}, \bibinfo
  {author} {\bibfnamefont {Ulrich}\ \bibnamefont {Schneider}}, \ and\ \bibinfo
  {author} {\bibfnamefont {Immanuel}\ \bibnamefont {Bloch}},\ }\bibfield
  {title} {\enquote {\bibinfo {title} {Observation of many-body localization of
  interacting fermions in a quasirandom optical lattice},}\ }\href {\doibase
  10.1126/science.aaa7432} {\bibfield  {journal} {\bibinfo  {journal}
  {Science}\ }\textbf {\bibinfo {volume} {349}},\ \bibinfo {pages} {842--845}
  (\bibinfo {year} {2015})}\BibitemShut {NoStop}%
\bibitem [{\citenamefont {von Oppen}\ \emph {et~al.}(1996)\citenamefont {von
  Oppen}, \citenamefont {Wettig},\ and\ \citenamefont
  {M\"uller}}]{vonoppen1996interaction}%
  \BibitemOpen
  \bibfield  {author} {\bibinfo {author} {\bibfnamefont {Felix}\ \bibnamefont
  {von Oppen}}, \bibinfo {author} {\bibfnamefont {Tilo}\ \bibnamefont
  {Wettig}}, \ and\ \bibinfo {author} {\bibfnamefont {Jochen}\ \bibnamefont
  {M\"uller}},\ }\bibfield  {title} {\enquote {\bibinfo {title}
  {Interaction-induced delocalization of two particles in a random potential:
  Scaling properties},}\ }\href {\doibase 10.1103/PhysRevLett.76.491}
  {\bibfield  {journal} {\bibinfo  {journal} {Phys. Rev. Lett.}\ }\textbf
  {\bibinfo {volume} {76}},\ \bibinfo {pages} {491--494} (\bibinfo {year}
  {1996})}\BibitemShut {NoStop}%
\bibitem [{\citenamefont {Arnoldi}(1951)}]{arnoldi1951the}%
  \BibitemOpen
  \bibfield  {author} {\bibinfo {author} {\bibfnamefont {W.~E.}\ \bibnamefont
  {Arnoldi}},\ }\bibfield  {title} {\enquote {\bibinfo {title} {The principle
  of minimized iteration in the solution of the matrix eigenvalue problem},}\
  }\href {\doibase 10.1090/qam/42792} {\bibfield  {journal} {\bibinfo
  {journal} {Quart. Appl. Math.}\ }\textbf {\bibinfo {volume} {9}},\ \bibinfo
  {pages} {17--29} (\bibinfo {year} {1951})}\BibitemShut {NoStop}%
\bibitem [{\citenamefont {Frahm}(2016)}]{frahm2016eigenfunction}%
  \BibitemOpen
  \bibfield  {author} {\bibinfo {author} {\bibfnamefont {Klaus~M.}\
  \bibnamefont {Frahm}},\ }\bibfield  {title} {\enquote {\bibinfo {title}
  {Eigenfunction structure and scaling of two interacting particles in the
  one-dimensional anderson model},}\ }\href {\doibase
  10.1140/epjb/e2016-70114-7} {\bibfield  {journal} {\bibinfo  {journal} {Eur.
  Phys. J. B}\ }\textbf {\bibinfo {volume} {89}},\ \bibinfo {pages} {115}
  (\bibinfo {year} {2016})}\BibitemShut {NoStop}%
\bibitem [{\citenamefont {Iyer}\ \emph {et~al.}(2013)\citenamefont {Iyer},
  \citenamefont {Oganesyan}, \citenamefont {Refael},\ and\ \citenamefont
  {Huse}}]{iyer2013many}%
  \BibitemOpen
  \bibfield  {author} {\bibinfo {author} {\bibfnamefont {Shankar}\ \bibnamefont
  {Iyer}}, \bibinfo {author} {\bibfnamefont {Vadim}\ \bibnamefont {Oganesyan}},
  \bibinfo {author} {\bibfnamefont {Gil}\ \bibnamefont {Refael}}, \ and\
  \bibinfo {author} {\bibfnamefont {David~A.}\ \bibnamefont {Huse}},\
  }\bibfield  {title} {\enquote {\bibinfo {title} {Many-body localization in a
  quasiperiodic system},}\ }\href {\doibase 10.1103/PhysRevB.87.134202}
  {\bibfield  {journal} {\bibinfo  {journal} {Phys. Rev. B}\ }\textbf {\bibinfo
  {volume} {87}},\ \bibinfo {pages} {134202} (\bibinfo {year}
  {2013})}\BibitemShut {NoStop}%
\end{thebibliography}%

\end{document}